%% file: fsqzcut.tex
\documentclass{article}
\usepackage{kmn,latexsym,subfigure,epsfig}
\newcommand{\mc}{\multicolumn}

\newcommand{\gta}{\stackrel{>}{_{\sim}}}

\input macro.tex

\def \cosone{$\Omega_{\rm M}=1$ and $\Omega_{\Lambda}=0$}
\def \costwo{$\Omega_{\rm M}=0.3$ and $\Omega_{\Lambda}=0.7$}

\begin{document}

\title[The flat-spectrum redshift cut-off]{On the 
redshift cut-off for flat-spectrum radio sources}
\author[Jarvis \& Rawlings]{Matt J.\,Jarvis\thanks{Email: mjj@astro.ox.ac.uk} \& Steve Rawlings \\
Astrophysics, Department of Physics, Keble Road, Oxford, OX1 3RH. \\
}
\maketitle

\begin{abstract}
We use data from the Parkes Half-Jansky Flat-Spectrum (PHJFS) sample
(Drinkwater et al. 1997) to constrain the cosmic evolution in the
co-moving space density $\rho$ of radio sources in the top decade of
the flat-spectrum radio luminosity function (RLF). A consistent
picture for the high-redshift evolution is achieved using both simple
parametric models, which are the first to allow for distributions in
both radio luminosity and spectral index, and variants of the $V /
V_{\rm max}$ test, some of which incorporate the effects of radio
spectral curvature.  For the most luminous flat-spectrum objects, the
PHJFS sample is extremely similar to that used by Shaver et al. (1996,
1998) to argue for an abrupt `redshift cut-off': a decrease by a
factor $\sim 30$ in $\rho$ between a peak redshift $z \sim 2.5$ and $z
\sim 5$. Our analysis finds that the observable co-moving volume is
too small to make definitive statements about any redshift cut-off for
the most luminous flat-spectrum sources, although both constant-$\rho$
(no cut-off) models and models with cut-offs as abrupt as those
envisaged by Shaver et al. are outside the 90\% confidence
region.  The inference that the decline in $\rho$ is most likely to be
gradual, by a factor $\sim 4$ between $z \sim 2.5$ and $z \sim 5$, is
in accordance with previous work on the RLF by Dunlop \& Peacock
(1990), but different to the abrupt decline favoured by studies of
optically-selected quasars. Dust obscuration provides one explanation
for this difference. We show that a significant fraction of the most
radio-luminous flat-spectrum objects are Giga-Hertz Peaked Spectrum
(GPS) rather than Doppler-Boosted (DB) sources, complicating any
interpretation of the redshift cut-off. Studies based on objects
extending into the next lower decade of the flat-spectrum RLF are
likely to be more fruitful but will require a separation of the GPS
and DB populations, careful radio selection and analysis of
$K$-corrections, and larger sky-area redshift surveys than those
currently available.
\end{abstract}

\section{Introduction}\label{sec:intro}
\begin{table*}[!ht]
\begin{center}
\begin{tabular}{c|c|c|c|c|c|c|c|c}
\hline\hline 
\mc{1}{c|}{Sample} & \mc{1}{c|}{Area/sr} & \mc{1}{c|}{Sample} &
\mc{1}{c|}{$0 < z \leq 1$} & \mc{1}{c|}{$1 < z \leq 2$} & \mc{1}{c|}{$2 < z \leq 3$} & \mc{1}{c|}{$3 < z \leq 4$} &  \mc{1}{c|}{$4 < z \leq 5$} \\ 
\hline\hline
PHJFS & 3.90 & $S_{2.7} > 0.5$\ Jy & 0 & 6 & 10 & 6 & 0 & \\
SH96$^{\dagger}$ & 4.0 & $S_{2.7} > 0.5$\ Jy & 0 & 7 & 12 & 6 & 0 &
\\
SH96$^{\dagger}$ & 3.8 & $S_{2.7} > 0.25$\ Jy & $-$ & $-$ & $-$ & $-$ & 0 & \\
PSR & 0.075 & $S_{2.7} > 0.1$\ Jy & 0 & 0 & 0 & 1 & 0 & \\
P85  & 0.58 & $S_{2.7} > 0.5$\ Jy & 0 & 0 &  3 & 2 & 0 & \\
PW81 & 4.05 & $S_{2.7} > 1.5$\ Jy & 0 & 1 &  2 & 0 & 0 & \\
WP85 & 9.81 & $S_{2.7} > 2.0$\ Jy & 0 & 6 &  7 & 0 & 0 & \\
\hline\hline
\end{tabular}
{\caption{\label{tab:samples} Redshift distributions of the most
luminous flat-spectrum sources, as defined in Sec.~\ref{sec:PHJFSsample},
in various samples selected at 2.7 GHz from the PHJFS, SH96, PSR, P85,
PW (Peacock \& Wall 1981) and WP85 (Wall \& Peacock 1985) samples.
The latter three samples were the bright samples used in the DP90
study and the PSR sample combines the Parkes Selected Regions to form
the faint sample used by DP90. The $\dagger$ symbol denotes the
samples where the spectral index selection criterion of
$\alpha_{2.7}^{5.0} \leq 0.4$ was used; the PHJFS and DP90 studies
adopted $\alpha_{2.7}^{5.0} \leq 0.5$ as the flat-spectrum
criterion. The `$-$' symbols indicate data yet to be published.}}
\end{center}
\end{table*}

A basic question in cosmological research is the 
redshift, or cosmic epoch, at which the first active galaxies
were born. Answering this question is important because active
galaxies can have significant impacts on the Universe, for example 
as a sources of photons for re-ionisation (Haiman \&
Loeb 1997) and as a source of entropy for the inter-galactic medium
(e.g. Valageas \& Silk 1999). Distant active
galaxies also provide a valuable probe
of the formation and early evolution of massive galaxies and their associated
dark-matter halos (e.g. Efstathiou \& Rees 1988).
The differential evolution of active galaxies and
the global star-formation rate has recently been recognised
as a powerful probe of the merger processes which underpin 
galaxy formation in hierarchical models of structure formation
(e.g. Percival \& Miller 1999; Cen 2000).

Investigations into the co-moving space density $\rho$ of
flat-spectrum radio sources selected at $2.7\,$GHz (Peacock 1985,
hereafter P85; Dunlop \& Peacock 1990, hereafter DP90), have found
evidence of a large increase in $\rho$ out to redshift $z \sim 2.5$,
and seemingly strong evidence for a decline in $\rho$ at higher
redshifts. This high-redshift decline, regardless of its magnitude,
has come to be known as the `redshift cut-off', a term which was first
introduced by Sandage (1972).  Its existence is often used to assert
that the $z \sim 2.5$ Universe corresponds to the epoch of maximum
quasar activity (e.g. Shaver et al. 1998; hereafter SH98).

Using a $2.7\,$GHz-selected sample, Shaver et al. (1996; hereafter
SH96) suggested a decline in $\rho$ of more than 1
dex between $z \sim 2.5$ and $z \sim 5$ ($\sim 1.5$ dex according to
Fig. 1 of SH98), whereas the models of DP90 suggested a much more
gradual decline, with behaviour not too far from a roughly constant
space density for the most luminous flat-spectrum
sources. Understanding this apparent contradiction provided the first
motivation for the work described in this paper.

DP90 were the first to suggest that observational data favours a
redshift cut-off in the steep- as well as the flat-spectrum
population, although their study was subject to a number of
uncertainties, most notably a reliance on photometric redshift
estimates for a large fraction of their high-redshift sources.  Wall
\& Jackson (1997) and Jackson \& Wall (1999) have developed a model
which explains the behaviour of both the flat- and steep-spectrum
populations using a unification scheme (e.g. Antonucci 1993) in which
the flat-spectrum sources are the Doppler-boosted (DB) products of a
parent steep-spectrum population in which a redshift cut-off appears
as a hard-wired feature. These studies paint a picture of radio source
evolution in which a high-redshift redshift cut-off is a natural
component (although not an essential component of the Wall \& Jackson
models; Wall, priv. comm.).

Over the last few years the Oxford group has led programmes aimed at
re-investigating this question using redshift surveys of samples
selected at low radio frequencies, and hence dominated by the
steep-spectrum population (e.g. Rawlings et al. 1998; Willott et
al. 1998; Jarvis et al. 1999; Willott et al. 2000; Jarvis et
al. 2000). These studies have yet to find any statistical
discrimination between models with constant $\rho$ at high-redshift,
and those with high-redshift cut-offs. Of course any high-redshift
decline in $\rho$ might be quite gradual or might be a strong function
of radio luminosity, so there is not necessarily any fundamental
disagreement between this work and claims of redshift cut-offs based
on much fainter samples of radio sources (e.g. Dunlop 1998).  With
these studies in mind, the second motivation of this paper was to
determine whether the cosmic evolution of the high-redshift
flat-spectrum population provide indirect evidence for a cut-off
which, in our opinion, has yet to be established beyond doubt in the
parent radio-luminous steep-spectrum population (but see DP90, Dunlop
1998).

As emphasised by P85, DP90, SH96 and others, the crucial advantage of
any radio-based work is that with sufficient optical follow-up, it can
be made free of optical selection effects, such as increasing dust
obscuration at high-redshift. It is chiefly for this reason that the
SH96 work is often highlighted as the most convincing evidence to date
for the existence of any redshift cut-off for the active galaxy
population. Indeed, the similarity of the decline in the $\rho$ of
radio sources to those of both optically-selected quasars and to the
global star-formation rate has been used (e.g Boyle \& Terlevich 1998;
Dunlop 1998; SH98; Wall 1998) to suggest a close link between the
triggering of starburst and AGN activities, and also to marginalise
the effects of dust obscuration on the optically-selected quasar
population.  However, recent results have cast severe doubts on the
veracity of this similarity if there is indeed an abrupt redshift
cut-off in the quasar population. The main reason for this is that the
most recent versions of the plot of global star-formation rate versus
redshift (e.g. Steidel et al. 1999) no longer feature any significant
high-redshift decline (c.f. Madau et al. 1996), although gentle
declines such as the decline found by DP90 are consistent, within the
uncertainties, with a roughly constant rate of star-formation at
high-redshift.  Recent papers (e.g. Cen 2000) have tended to emphasise
and model the difference between the roughly constant star-formation
rate from $z \sim 2$ to $z \sim 4$ and the abrupt redshift cut-off inferred
for optically selected quasars.  A desire to understand whether this difference is driven
by increased dust obscuration at high-redshift provided the third
motivation for this paper.
\begin{center}
\begin{figure*}[!ht]
{\hbox to 0.9\textwidth{ \null\null \epsfxsize=0.9\textwidth
\epsfbox{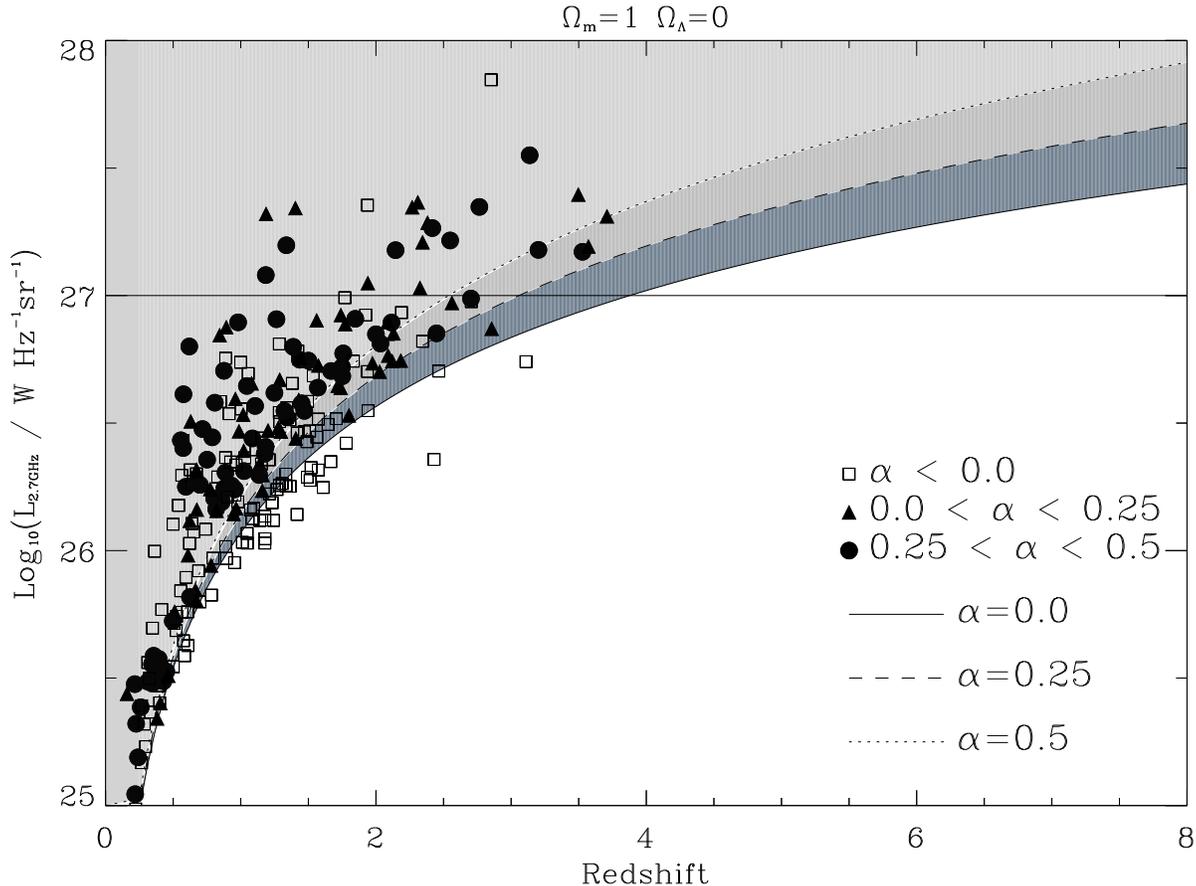} }}
{\caption{\label{fig:pz_1} The rest-frame 2.7\,GHz radio luminosity
$L_{2.7}$ versus redshift $z$ plane for the PHJFS sample
for cosmology I. The horizontal solid line shows the lower luminosity
limit in cosmology I. The solid curve corresponds to the lower limit
in luminosity for sources with $\alpha = 0.0$, the dotted line for
sources with $\alpha=0.25$ and the dashed line for sources with
$\alpha=0.5$. The shaded regions show where sources at the flux-density limit ($S_{2.7} \geq 0.5$\,Jy) are permitted to lie. The filled
circles represent sources from the PHJFS with radio spectral indices
in the range $0.25 < \alpha < 0.5$;
filled triangles, $0.0 < \alpha < 0.25$; and open squares, $\alpha <
0.0$, i.e. the inverted-spectrum sources.}}
\end{figure*}
\end{center}

The question of the quasar redshift cut-off is just beginning to be addressed
by X-ray surveys, and we can expect rapid advances in this field
with the advent of surveys made with Chandra and XMM-Newton.
The most recent evaluations of the high-redshift evolution of quasars, 
based on soft X-ray selected samples (Miyaji, Hasinger \& Schmidt 2000),
find no firm evidence for an abrupt cut-off. Comparison of the space
density at high-redshift from radio and X-ray measurements provided the
fourth motivation for this paper.

P85 and SH96 have highlighted the potential problem of spectral
curvature and its effect on obtaining reliable $K-$corrections for
distant flat-spectrum radio sources: a concave spectral shape means
that redshifting produces a systematic increase in the spectral index
between two fixed observed frequencies as redshift increases.  SH96 used the
Gear et al. (1994) study to determine at what redshift $z$ a flat-spectrum
source would have an observed spectral index $\alpha_{2.7}^{5.0} >
0.4$\footnote{We use the spectral index convention $S_{\nu} \propto
\nu^{-\alpha}$, where $S_{\nu}$ is the flux-density at the observing
frequency $\nu$.}, finding $z \gta 10$.  However, Gear et al. observed
only DB objects (BL Lacs and OVV quasars), and it was not clear to us
that these are necessarily representative of the flat-spectrum
population at the highest radio luminosities.  Pursuing this worry
provided a fifth and final motivation for this paper.

In Sec.~\ref{sec:PHJFSsample} we describe the PHJFS sample and how it
relates to the studies of P85, DP90 and SH96. In Sec.~\ref{sec:imp_spix} we
highlight the important r\^{o}le of the distribution in spectral index to
any investigation of the RLF, and of the dangers of using a binned
estimation of the RLF. In Sec.~\ref{sec:imp_curv} we emphasise
the importance of spectral curvature. Sec.~\ref{sec:modelling} outlines a
simple parametric modelling procedure which incorporates a
distribution in spectral index and presents the results of our
modelling. In Sec.~\ref{sec:VVmax} we use variants of the $V / V_{\rm
max}$ statistic to further investigate the high-redshift space density. In
Sec.~\ref{sec:cons_uncert} we attempt to constrain the uncertainties of
any high-redshift decline. The implications of our results with particular
reference to the five motivations outlined in this Introduction are
discussed in
Secs.~\ref{sec:discussion} and~\ref{sec:cosmo_context}. We review
prospects of constraining the space density of flat-spectrum quasars
with future redshift surveys in Sec.~\ref{sec:estab_cutoff}.

We take $H_{\circ} = 50\,{\rm km\,s}^{-1}\,{\rm Mpc}^{-1}$ and use two
cosmological models: cosmology I is defined by the dimensionless
parameters \cosone; cosmology II by \costwo. All radio luminosities quoted
are measured in units of W\,Hz$^{-1}$\,sr$^{-1}$.

\section{The Parkes Half-Jansky Flat-Spectrum sample}\label{sec:PHJFSsample}
The Parkes Half-Jansky Flat-Spectrum sample (PHJFS) contains 323
sources selected at $2.7\;$GHz with a flux-density $S_{2.7} >
0.5\;$Jy, and a spectral index measured between 2.7 and 5.0$\;$GHz,
$\alpha_{2.7}^{5.0} < 0.5$. The survey covers a sky area of 3.90
steradians over all right ascension and declinations $-45^{\circ} < \delta(1950) <
+10^{\circ}$ excluding galactic latitudes $|b| < 20^{\circ}$. Most (281) of
these sources now have spectroscopic redshifts, a completeness of
$\approx 87$ percent. We will assume throughout that no significant
biases are introduced by this small redshift incompleteness. 

We will focus our investigation on the most radio-luminous sources.
For cosmology I, we consider the objects with $\log_{10}(L_{2.7}) \geq
27.0$ which isolates approximately the top-decade in luminosity, and
is also a very similar criterion to that used by SH96 in their
analysis. We also model the RLF in cosmology II with a higher
luminosity limit of $\log_{10}(L_{2.7}) \geq 27.3$ which corresponds
to the same number of sources present in the analysis for cosmology I,
and also corresponds roughly to the top-decade of the RLF.  The
redshift distributions of the most luminous sources in the PHJFS and
of the sample of SH96 are very similar (see Table~\ref{tab:samples})
and a 1-D Kolmogorov-Smirnov test gives a probability of $P_{KS} =
0.99$ suggesting that the redshift distributions are statistically
indistinguishable.

It might seem bizarre to concentrate on such a small subset of the
total PHJFS sample for our statistical analysis, but the reason for
this is that, following SH96, we are seeking {\em direct} evidence for
the decline in the quasar population at high-redshifts. Less radio
luminous objects are simply not detectable at high-redshifts in
relatively bright samples like the PHJFS. Fig.~\ref{fig:pz_1} shows
the radio luminosity - redshift ($L_{2.7}-z$) plane for the PHJFS
sample including the loci of its $S_{2.7} \geq 0.5\,$Jy flux-density
limit at three different values of $\alpha$. For an $\alpha \sim 0$
source the maximum observable redshift of a source at the PHJFS
flux-density limit and our adopted luminosity limit is four, and this
redshift limit drops precipitously as the critical luminosity is
lowered.

Obviously many of the PHJFS sources will also be present in the SH96
sample, due to the overlap in sky area and the similar flux-density
limit. Indeed $\approx 65 \%$ of the sky-area covered by the PHJFS is
also covered by the $S_{2.7} > 0.5$\ Jy sample of SH96. Therefore one
would expect at least $\approx 14$ of the most luminous sources in the
PHJFS to be included in the SH96 sample. Although one might expect to
find a few more sources in total in the SH96 sample due to the
incompleteness of the PHJFS.  The overlap with the samples used by
DP90 will be less ($\sim 50$ \%) due to their use of the comparatively
deep $S_{2.7} > 0.1$\ Jy Parkes selected region (1 source in common),
and poor sky overlap with some of the brighter samples. Of the 22
PHJFS objects considered here, 7 sources are in common with the sample
of WP85, 3 sources with P85, and the $S_{2.7} > 1.5$\ Jy sample has no
overlap with the PHFJS as it only covers declinations $\delta >
10^{\circ}$.

\section{The importance of modelling the spectral index distribution and the RLF}\label{sec:imp_spix}
\begin{figure*}[!ht]
{\hbox to 0.45\textwidth{ \null\null \epsfxsize=0.45\textwidth
\epsfbox{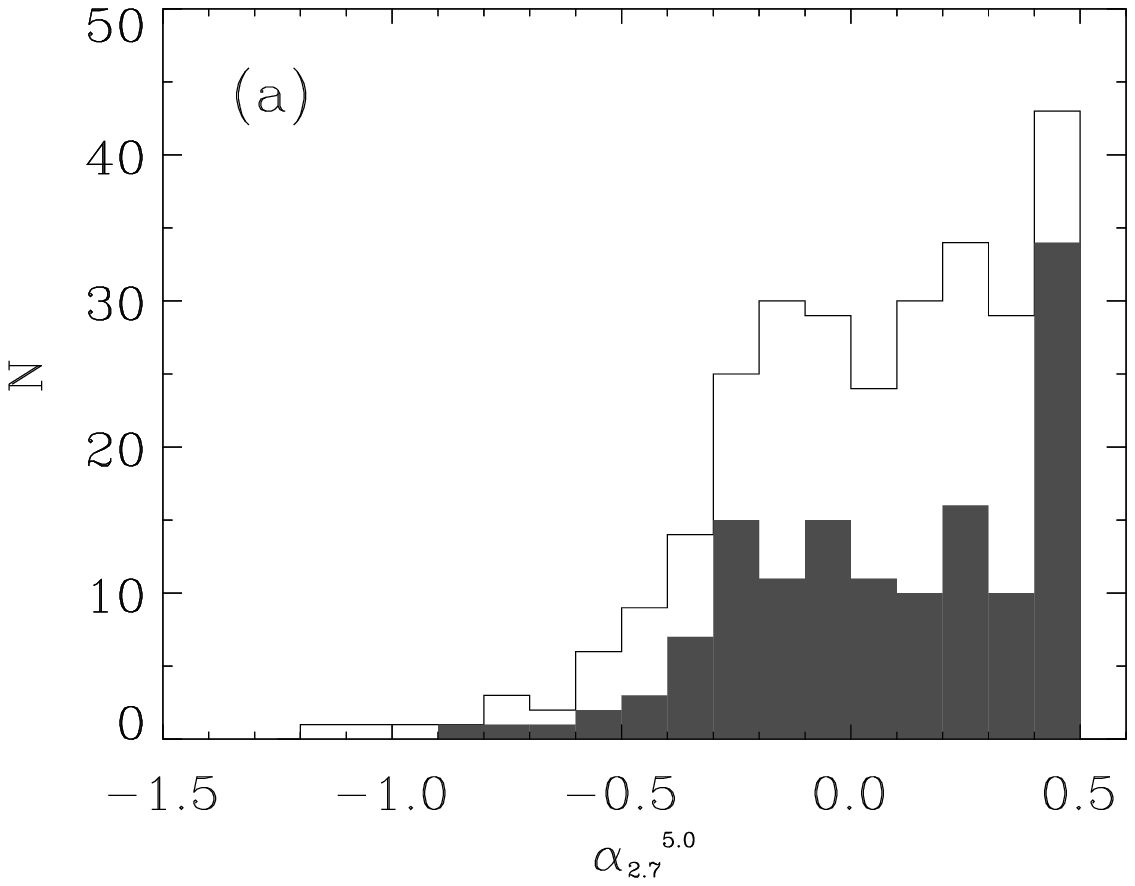}
\epsfxsize=0.45\textwidth
\epsfbox{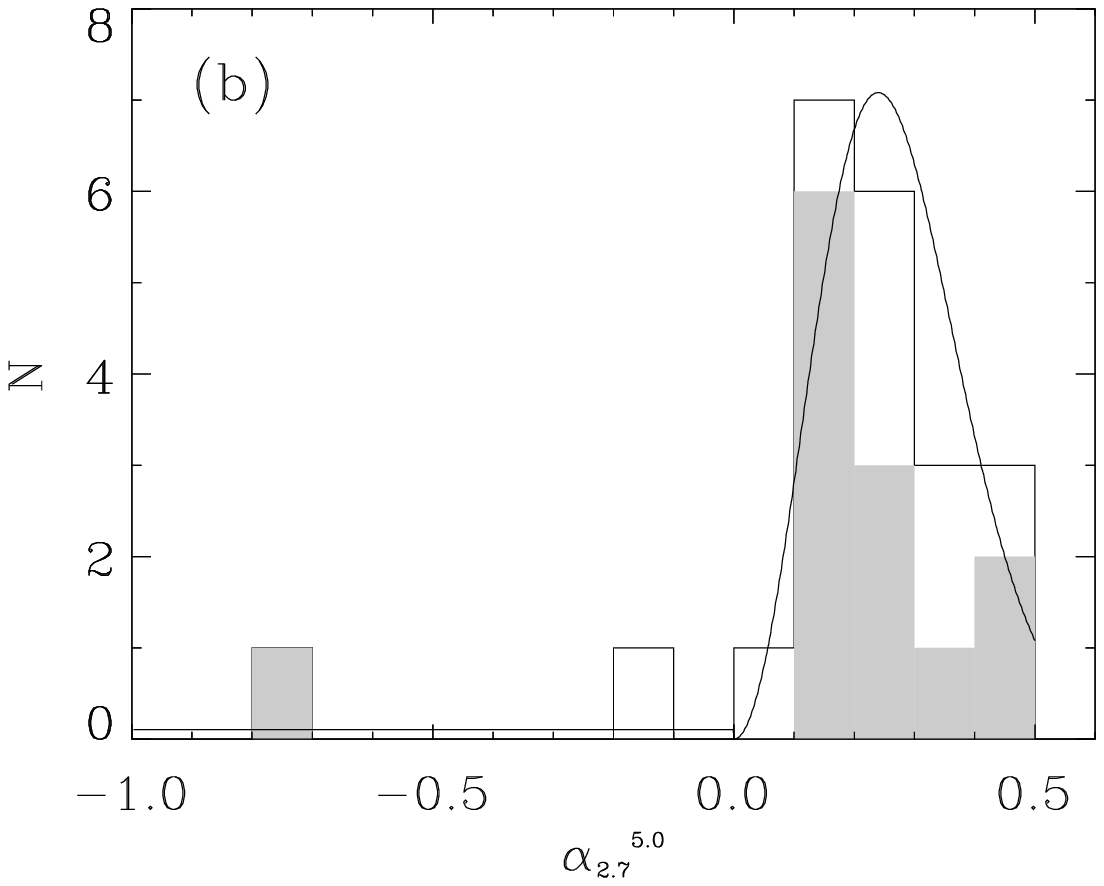} }}
{\caption{\label{fig:alpha_3} (a) Distribution in observed spectral
index between 2.7 and 5\,GHz ($\alpha_{2.7}^{5}$) for the whole of the
PHJFS sample with spectroscopic redshifts. The dark-shaded region
shows the distribution for sources with $z < 1$. (b) Distribution in
$\alpha$ for sources from the PHJFS with $\log_{10} (L_{2.7}) \geq
27.0$ for cosmology I. The curve shows the functional form used to
parameterise the distribution in $\alpha$ with the fitted values from
model A (Sec.~\ref{sec:modelling}) with the light-shaded region shows the
sources with $z < 2.5$.}}
\end{figure*}
\begin{figure}[!ht]
{\hbox to 0.45\textwidth{ \null\null \epsfxsize=0.45\textwidth
\epsfbox{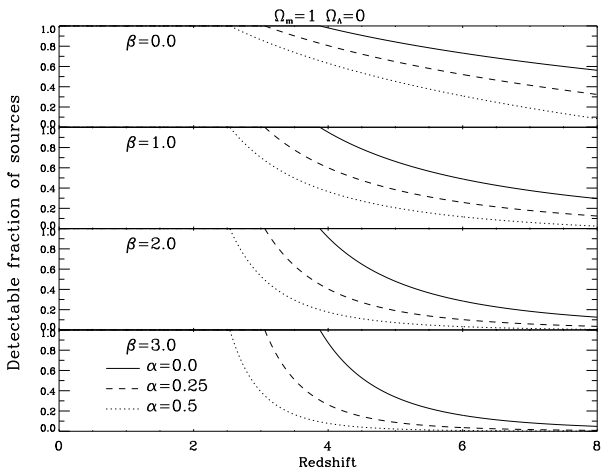} }}
{\caption{\label{fig:perc_2} The fraction of luminous
($\log_{10} (L_{2.7}) > 27$) sources on our light cone with $S_{2.7} =
0.5\,$Jy and spectral index $\alpha$ detectable in a flux-density limited
sample of $S_{2.7} \geq 0.5\,$Jy, weighted according to various
power-law RLFs (each with index $\beta$, see eqn.~\ref{eqn:rlfbeta}) for
cosmology I.}}
\end{figure}

The name `flat-spectrum quasar' unfortunately carries with it the
implication that $\alpha \sim 0$ so that the flux-density, and hence
the luminosity, is more-or-less independent of frequency. One might
naively expect that $K-$corrections are therefore unnecessary, or of
marginal significance. As quantified by P85, even with the smooth and
fairly flat spectra of interest, this is not the case.

There are a finite number of radio-luminous flat-spectrum sources
observable on our light cone. The flux-density limit of a survey means
that only a fraction of these objects will make it into the survey
once the redshift exceeds a critical value given by the intersection
of the horizontal line in Fig.~\ref{fig:pz_1} and the relevant flux-density limit.  Fig.~\ref{fig:perc_2} illustrates this fraction as a
function of redshift assuming a power-law RLF of steepness $\beta$
(see eqn.~\ref{eqn:rlfbeta}) and a mean radio spectral index $\alpha$.
For a given $\beta$ at any given redshift above $z \sim 2.5$, the
observable fraction of the most luminous sources is a strong function
of spectral index because of the larger cosmological volume available
to the flatter-spectrum sources on our light cone (see
Fig.~\ref{fig:pz_1}). A second more subtle effect concerns $\beta$: at
fixed $\alpha$ and $z$, again assumed to be above $z \sim 2.5$, the
fraction of sources in the survey drops dramatically with $\beta$
because more luminous sources can be seen over larger cosmological
volumes, and $\beta$ determines the relative numbers of sources as a
function of luminosity. In the extreme case of a very steep RLF
($\beta \sim 3$), and $\alpha \sim 0.5$, for example, there is
effectively zero available volume on our light cone in which to detect
the most radio luminous flat-spectrum sources beyond $z \sim 4$. The
lack of very high-redshift quasars in a sample might be telling us
more about the lack of observable volume than about an intrinsic lack
of objects.

The analysis of the most radio luminous population by SH96 and SH98
adopted the median spectral index of their large sample, namely
$\alpha=0$ (Shaver, priv. comm.). However, focusing on the
most-luminous sources in the PHJFS we see from Fig.~\ref{fig:pz_1},
and clearer still in the histograms plotted in Fig.~\ref{fig:alpha_3},
that their spectral index distribution (Fig.~\ref{fig:alpha_3}b) is
dissimilar to the distribution for the whole sample
(Fig.~\ref{fig:alpha_3}a): Fig.~\ref{fig:alpha_3}a is a broad
distribution with a mean spectral index of $0.04 (\pm 3$\%) whereas
Fig.~\ref{fig:alpha_3}b has a sharp peak with a significantly steeper
mean spectral index of $0.19 (\pm 5$\%); there is still a tail of
sources with inverted spectra in Fig.~\ref{fig:alpha_3}b but it
includes only 9 per cent of the population whereas, in
Fig.~\ref{fig:alpha_3}a, 43 per cent are inverted. A 1-D
Kolmogorov-Smirnov test gives a probability $P_{KS} \sim 10^{-3}$ that
the two $\alpha$ distributions shown in Figs.~\ref{fig:alpha_3}a
and~\ref{fig:alpha_3}b are different. Fig.~\ref{fig:alpha_3}a, shows
the distribution in spectral index for sources with $z < 1$ (roughly
half the sources) which, being similar to the distribution as a whole
($P_{\mathrm {KS}} = 0.41$), suggests no gross dependence of the
spectral index distribution on redshift. Similarly
Fig.~\ref{fig:alpha_3}b shows the distribution of the most luminous
sources with $z < 2.5$ (again roughly half the sources) and once more
the distribution is consistent with the most luminous sources as a
whole ($P_{\mathrm {KS}} = 0.99$), and no gross redshift-dependent
effect is apparent.  Although redshift effects may play an important
r\^{o}le in determining the spectral index distribution of the sample, and
the analysis presented here falls far short of a thorough
investigation of the various inter-correlations, it is hard to escape
the conclusion that the correlation between luminosity and spectral
index is the dominant factor in skewing the distribution in $\alpha$
for the most radio luminous sources. We will suggest a probable cause
for this effect in Sec.~\ref{sec:linking_steep}.  Because the mean of the
distribution in Fig.~\ref{fig:alpha_3}b, i.e. $\alpha = 0.19 (\pm
5\%$), is significantly different to the $\alpha=0$ adopted by SH96,
there will be a significant reduction in the observable co-moving
volume of the most radio luminous flat-spectrum sources at
high-redshift (c.f. Fig.~\ref{fig:perc_2}).

The analyses of SH96 and SH98 employed a binning method for their
study of the most luminous flat-spectrum radio sources.  The finite
size of the bins in radio luminosity can also lead to systematic
effects. Consider the extreme example of analysing the $27.0 \leq
\log_{10} (L_{2.7}) \leq 28.0$ region of the PHJFS sample as a single
luminosity bin. Assigning $\log_{10} (L_{2.7}) = 27.5$ as the
characteristic luminosity of this bin would mean that the observable
volume would tend to be greatly over-estimated: the steepness of the
RLF means that most of the sources in the relevant luminosity bin
could only be observed over a much lower cosmological volume. This
would lead to an estimate of space density that was systematically
biased to low values. The influence of $\beta$ on the
fraction of luminous objects observable has already been shown in
Fig.~\ref{fig:perc_2}.

To summarise, we have identified in this section two possible systematic
effects in the SH96 and SH98 analyses which would tend to bias their
derived high-redshift space densities to low values. 
To quantify these effects we use parametric models for the RLF
(Sec.~\ref{sec:modelling}) which include a
distribution in $\alpha$, and which take account of the steepness of
the RLF.

\section{The effects of spectral curvature}\label{sec:imp_curv}
\begin{figure*}
{\hbox to 0.9\textwidth{ \null\null \epsfxsize=0.8\textwidth
\epsfbox{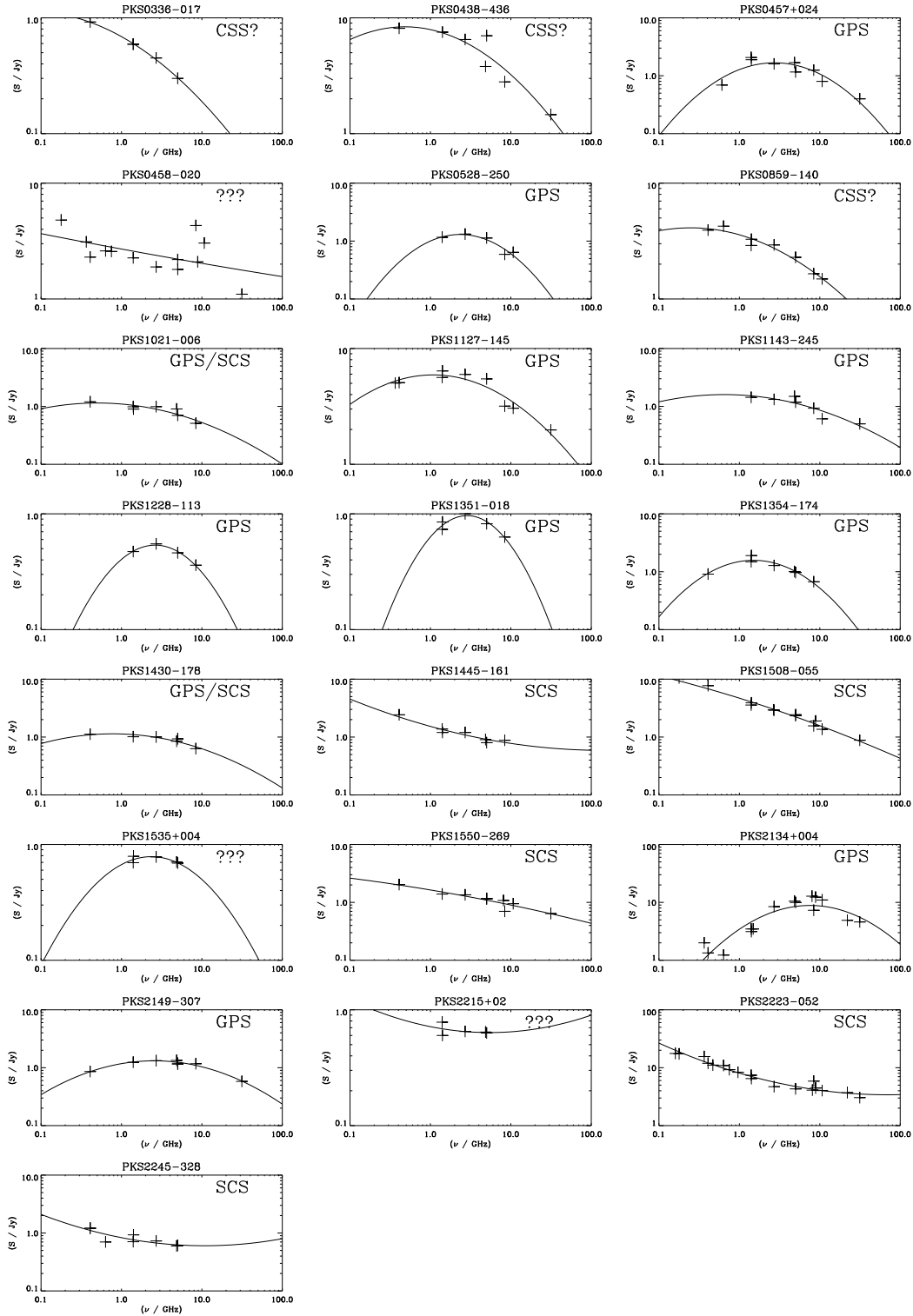}
}}
{\caption{\label{fig:radspec_6} Radio spectra of the most luminous
sources from the PHJFS. 
We have roughly classified each spectrum according to the following scheme:
GPS = Giga-Hertz Peaked Spectrum; CSS? =
possible Compact Steep Spectrum, spectrum peaking below
0.5\,GHz; SCS = Straight or Concave Spectrum;
GPS/SCS = GPS or SCS (more observations needed to clarify); and
??? = not classified due to insufficient data.
The curves show the fits described in Sec.~\ref{sec:imp_curv}.
}}
\end{figure*}
The radio spectra of the most luminous PHJFS sources
(Fig.~\ref{fig:radspec_6}) show that, as suggested by Savage \&
Peterson (1983), P85 and SH96, spectral curvature needs to be
considered. In this short section we quantify how this might
influence searches for very high-redshift quasars at high ($2.7$\,GHz)
radio frequencies.

By fitting the spectra in Fig.~\ref{fig:radspec_6} with a polynomial
of the form $y = \log_{10} S_{\nu} = \sum_{i=0}^{2} a_{i}x^{i}$, where
$x = \log_{10} (\nu/{\rm GHz})$, we find that $\approx 70\%$ of the
spectra become steeper with increasing frequency with mean values of $a_{1} =
0.07 \pm 0.08$ and $a_{2} = -0.29 \pm 0.06$. If we now use these mean
values as representative of a source at $z \simeq 2.5$ then we find
that it would have an observed spectral index between 2.7 and 5.0\,GHz
$\alpha_{2.7}^{5.0} \simeq 0.26$ which is similar to the mean of our
sample just as one would expect. If we now shift this source out to $z
= 5$ then it would have steepened to $\alpha_{2.7}^{5.0} \simeq
0.4$. The mean spectral index will therefore increase systematically
with redshift when the effects of spectral curvature are considered
and, as we will show in Sec.~\ref{sec:estab_cutoff}, even small systematic
shifts can produce significant effects. Sources at $z \geq 5$ in a
flux-density limited sample are likely to have significantly reduced observable
volumes (see Fig.~\ref{fig:perc_2}) as a result of spectral curvature, and also
in some cases to become so steep that they fail the filtering criteria
of the survey: current searches for high-redshift flat-spectrum
quasars (e.g. SH96) do not include sources whose observed spectra are
steeper than a critical value around 0.4.

\section{Parametric RLF modelling}\label{sec:modelling}
\subsection{Method}\label{sec:modelmethod}
Parametric models provide an effective method of extracting as much
information as possible from a small dataset, and according to Occam's
razor we need to favour the simplest model consistent with the data;
in practice, this requires us to restrict the models to as few free
parameters as possible.  An alternative approach is to use a
`free-form' fit (Peacock \& Gull 1981; P85; DP90) which has the
advantage of making more general assumptions about the functional
forms of the model RLFs, but which with the small dataset of interest
here would require the use of such low-order `free-form' polynomial
expansions that it would end up using functional forms which are very
similar to those we explicitly consider.

To investigate the RLF of the most-luminous flat-spectrum
quasars we considered six models with common parameterisations for the
radio luminosity dependence and spectral index distribution, but
differing parameterisations for the redshift evolution. That is we
look for a separable distribution function of the form
\begin{equation}
\rho(L_{2.7},z,\alpha) = \rho_{\circ} \times \rho_{L}(L_{2.7}) \times
\rho_{X}(z_{N}) \times \rho_{\alpha}(\alpha) , 
\end{equation}
where the normalising factor $\rho_{\circ}$ is a free parameter 
measured in units of Mpc$^{-3}$ and $\rho_{L} (L_{2.7})$, $\rho_{X} (z)$ and
$\rho_{\alpha} (\alpha)$ are dimensionless distribution functions per $(\Delta
\log_{10}L_{2.7})$, per $(\Delta z)$, and per $(\Delta \alpha)$ respectively.
Our treatment of a distribution in $\alpha$ means that
our definition of $\rho$ differs from that employed by, for example,
DP90.

It is highly improbable that the true distribution function is
separable since there are likely to be cross-correlations between
$L_{2.7}$, $z$ and $\alpha$ from both physical and $K$-correction
effects.  However, by confining our attention to a narrow range in
$L_{2.7}$, the influence of such cross-correlations are
minimised, and our assumed separable form is likely to be an adequate
approximation. The alternative of encoding correlations into the
functional form would introduce too many additional free parameters
for the small dataset under study.  We return to the possible
influence of $z$--$\alpha$ correlations in Sec.~\ref{sec:estab_cutoff}
when we consider searches for flat-spectrum quasars at $z > 5$.

We use a single power-law to parameterise $\rho(L_{2.7})$, i.e.
\begin{equation}\label{eqn:rlfbeta}
\rho_{L}(L_{2.7}) = \left (\frac{L_{2.7}}{L_{\circ}}\right)^{-\beta} , 
\end{equation}
where $\beta$ is a dimensionless free parameter, $L_{2.7}$ is the
rest-frame 2.7\,GHz luminosity and $L_{\circ}$ is a normalising
luminosity fixed at the lower luminosity limit of the sample of the
most luminous sources (different for each cosmology).  We parameterise the
distribution of spectral indices with
\begin{equation}
\label{eqn:alpha_dist}
\rho_{\alpha}(\alpha) =  \left\{ \begin{array} {l@{\quad:\quad}l}
\alpha^{2}\exp(-\gamma \alpha^{2}) & \alpha > 0.0 \\ \epsilon & \alpha
\leq 0.0 , \end{array} \right. 
\end{equation}
where $\gamma$ and $\epsilon$ are free parameters in the model
fitting.  We now describe the redshift distributions for the six
models. Model A is parameterised by a single Gaussian distribution in
redshift which (at least in cosmology I) is consistent with the shape
of the evolution in co-moving space density of the most luminous
flat-spectrum sources suggested by SH96, and also with some
derivations of the evolution of optically-selected quasars (see
Fig.~\ref{fig:perc_2} of SH96). The form used forces a decline at high
redshifts, i.e.
\begin{equation}
\rho_{A}(z) = \exp \left \{-\frac{1}{2} \left(
\frac{z-z_{\circ}}{z_{1}} \right)^{2} \right \} ,
\end{equation}
where $z_{\circ}$ is the redshift of the Gaussian peak and $z_{1}$ is
the characteristic width of the Gaussian. Note that an undesirable
feature of this functional form is that it
enforces a symmetry about $z_{0}$ which couples the decay rates at 
low- and high-redshift for which there is no physical justification.

Model B is parameterised by a Gaussian which becomes constant beyond
its peak, i.e.
\begin{equation}
\rho_{B}(z) = \left\{ \begin{array} {l@{\quad:\quad}l} \exp \left \{-\frac{1}{2} \left(
\frac{z-z_{\circ}}{z_{1}} \right)^{2} \right \} & z \leq z_{\circ} \\
1.0 & z > z_{\circ} , \end{array} \right.
\end{equation}
where $z_{\circ}$ and $z_{1}$ are as previously defined. 
These two forms have also been used by Willott et al. (1998)
in their study of the RLF of steep-spectrum radio quasars.

Models C-E use cut-offs at high- and low-redshift set to be equal to relevant
dynamical time-scales. At low-redshifts we use 
$t_{clus} = 4 \times 10^{9}$ yr, the
dynamical time-scale for a massive ($\sim 10^{15}$\,M$_{\odot}$) cluster, as the
width of a Gaussian in redshift space where the activity has fallen by
a factor of $\sim 100$ from the fitted peak.
The rationale for this is that the cause of the low-redshift
decline in $\rho$ seems to be linked to the virialization
of rich clusters (e.g. Ellingson, Green \& Yee 1991, Rees 1995) since 
galaxy mergers,
the probable trigger mechanism for powerful radio activity,
are suppressed in such environments.
This assumption is somewhat arbitrary, but serves to eliminate
fitting problems with setting the model $\rho$ precisely to zero at
some redshift, and also clearly decouples the low-redshift
evolution from the high-redshift evolution.

Model C parameterises the low-redshift cut-off as just described, and fits
the high-redshift evolution with a half Gaussian
above a fitted peak, i.e.
\begin{equation}
\rho_{C}(z) = \left\{ \begin{array} {l@{\quad:\quad}l} \exp \left
\{-\frac{1}{2} {\left(\frac{z-z_{\circ}}{z_{clus}/2.58}\right)^{2}}
\right \} & z \leq z_{\circ} \\ \exp \left
\{-\frac{1}{2} {\left(\frac{z-z_{\circ}}{z_{1}}\right)^{2}} \right \}
 & z > z_{\circ} , \end{array} \right. 
\end{equation} 
where $z_{\circ}$ and $z_{1}$ are again the peak and width of the
Gaussian; $z_{clus}$ corresponds to a time $t_{clus}$ after the peak
where the activity has decreased by $\approx 99\%$. The factor of 2.58
reduces the width of the Gaussian to a value which corresponds to the
inclusion of 99 per cent of the area of the full Gaussian for the given
time-scale in redshift space.

Model D again uses $t_{clus}$ to set the rate of
the low-redshift decline but forces the distribution
to a constant above a fitted $z_{\circ}$, i.e.
\begin{equation}
\rho_{D}(z) = \left\{ \begin{array} {l@{\quad:\quad}l} \exp \left
\{-\frac{1}{2} {\left(\frac{z-z_{\circ}}{z_{clus}/2.58}\right)^{2}}
\right \} & z \leq z_{\circ} \\ 1.0
 & z > z_{\circ} , \end{array} \right. 
\end{equation} 
where the symbols are consistent with those used in Model C. Note that
this model uses only five free parameters. 

For model E we use $t_{gal} = 3 \times 10^{8}$ yr, the dynamical
time-scale of a massive ($\sim 10^{12}$\,M$_{\odot}$) galaxy, to set the
decay rate of the high-redshift cut-off.  Again this choice is somewhat
arbitrary, but the rationale is that the cut-off at high-redshift is
probably linked to the formation of the first massive galaxies
(e.g. Rees 1995), and their dynamical time-scale sets the fastest rate
at which the cut-off can decline.  Below the fitted peak $z_{\circ}$
this model is again constrained by $t_{clus}$ and above $z_{\circ}$
this model becomes constant until it reaches $z_{1}$ where it declines
as a Gaussian, i.e.
\begin{equation}
\rho_{E}(z) = \left\{ \begin{array} {l@{\quad:\quad}l} \exp \left
\{-\frac{1}{2} {\left(\frac{z-z_{\circ}}{z_{clus}/2.58}\right)^{2}}
\right \} & z \leq z_{\circ} \\ 1.0 & z_{\circ} < z < z_{1} \\ \exp
\left \{-\frac{1}{2}
{\left(\frac{z-z_{\circ}}{z_{gal}/2.58}\right)^{2}} \right \} & z \geq
z_{1} , \end{array} \right. 
\end{equation} 
where $z_{gal}$ is the width in redshift space which corresponds to
$t_{gal}$ beyond the peak where the activity has decreased by a factor of $\approx \,99$.

Finally model F fixes the space density to be constant with redshift 
over all redshifts, i.e.
\begin{equation}
\rho_{F}(z) = 1.0 .
\end{equation}
This model will be used to illustrate the problem of
dealing with small number statistics at low- and high-redshift where
the observable co-moving volume is low.
Note that this model has only four free parameters.

We used the maximum likelihood method of Marshall et
al. (1983) to find best-fit parameters for all six models. If we define
$S$ as $-2 \, {\rm ln}\mathcal{L}$, where $\mathcal{L}$ is the likelihood
function, then by minimising $S$ we find the best-fit values for the
free parameters. $S$ is given by,
\begin{eqnarray}
\label{eqn:S}
S=-2\sum^{N}_{i=1}\ln [\rho(L_{{2.7}_{i}},z_{i},\alpha_{i}] \nonumber \\
\;\;\;\;\;\;\;\;\; +2\int\!\!\!\int\!\!\!\int
\rho(L_{2.7},z,\alpha)\Omega(L_{2.7},z,\alpha) \nonumber \\
\times \frac{{\rm d} V}{{\rm d}z} {\rm d}z \,{\rm
d}(\log_{10}L_{2.7})\, {\rm d}\alpha ,
\end{eqnarray}
where $\rho(L_{2.7},z,\alpha)$ is the model distribution,
$\Omega(L_{2.7},z,\alpha)$ is the sky area available (in sr)
for samples of a given flux-density and spectral index cut-off, and
$({\rm d} V/ {\rm d} z)$ is the differential co-moving volume element
per steradian. The first term is simply the sum over the $N$ sources in the
defined sample. The second term is the integrand of the model
distribution and should yield $\approx 2 N$ for good fits.  A downhill
simplex routine was used to minimise eqn.~\ref{eqn:S} to find the
best-fit parameters. The errors associated with these parameters were
found by numerically calculating the components of the {\sl Hessian}
matrix ($\nabla\nabla S$) at the location of the minimum, inverting
this matrix to obtain the covariance matrix $([\sigma^{2}]_{ij} =
2[(\nabla\nabla S)^{-1}]_{ij})$ and taking the $1\sigma$
errors as given by the square root of the diagonal elements of this
matrix (e.g. Sivia 1996).

To determine whether the best-fit models are reasonable fits to the data we 
used 1- and 2-D Kolmogorov-Smirnov (KS) tests (Peacock 1983)
on projections of the data. From these we determined the probability
$P_{KS}$ that the model distribution is a fair representation of the
data; note that in cases where the probability $P_{KS} > 0.2$,
the model and data are statistically indistinguishable, 
and that higher values of $P_{KS}$ are not necessarily suggestive
of better fits (see Press et al. 1992). 

To obtain relative probabilities for each model we used the procedure
set out in Sivia (1996). Briefly, for the models under consideration
the ratio of posterior probabilities of model X relative to model C is given approximately by
\begin{equation}\label{eqn:rel_prob}
P_{\rm R} = \frac{P({\rm X}\, |\, {\rm data})}{P({\rm C}\, |\, {\rm data})}
= \frac{e^{\frac{-S_{X | {\rm {min}}}}{2}}\sqrt{{\rm  Det}(\nabla\nabla
S_{C}|_{S_{\rm min}})}}{e^{\frac{-S_{C | {\rm {min}}}}{2}}\sqrt{{\rm  Det}(\nabla\nabla
S_{X}|_{S_{\rm min}})}} \times {\mathcal F},
\end{equation}
where ${\rm Det}(\nabla\nabla S_{X}|_{S_{\rm min}})$ is the
determinant of the Hessian matrix for model X, evaluated at $S =
S_{\rm min}$ and prior ranges for free parameters
common to all models (e.g. those of $\log_{10}\rho_{\circ}$, $\beta$,
$\gamma$ and $\epsilon$) have been cancelled. ${\mathcal F}$ is a
factor which compensates for the varying number of parameters between
models, and for model F, for example is
\begin{displaymath}
{\mathcal F} = (4\pi)^{\frac{N-6}{2}} \Delta z_{\circ} \Delta z_{1},
\end{displaymath}
where $N$ is the number of free parameters (i.e. 4 for model F) and
$\Delta z_{\circ} \Delta z_{1}$ is the multiple of the prior ranges
for any additional parameters in model C (with respect to model F
\footnote{Note that in a comparison of model D with model C this multiple of
the prior ranges would become just $\Delta z_{1}$.}). In all cases
considered here, this additional factor ${\mathcal F}$ is of order
unity and may be ignored.
\subsection{Results}\label{sec:modelresults}
\begin{table*}
\scriptsize
\begin{center}
\begin{tabular}{c|c|c|c|c|c|c|c|c|c|c|l}
\hline\hline \mc{1}{c|}{Model} & \mc{1}{c|}{Cos} & \mc{1}{c|}{$N$}
& \mc{1}{c|}{$\log_{10}\rho_{\circ}$} &
\mc{1}{c|}{$\beta$} & \mc{1}{c|}{$z_{\circ}$} & \mc{1}{c|}{$z_{1}$} &
 \mc{1}{c|}{$\gamma$} &  \mc{1}{c|}{$\log_{10} \epsilon$}
& \mc{1}{c|}{$S_{\rm {min}}$} & \mc{1}{c|}{$\ln{\rm Det}(\nabla\nabla S)$} &
\mc{1}{c|}{$P_{\rm R}$} \\ 
\hline\hline
A & I & 6 & $-7.29$ $\pm$ 0.18 & 1.30 $\pm$ 0.39 & 2.58 $\pm$ 0.22 &
0.87 $\pm$ 0.18 & 17.5 $\pm$ 3.7 & $-3.26$ $\pm$ 0.35 & 1037.67 & 29.38 & 0.85 \\
B & I & 6 & $-7.42$ $\pm$ 0.17 & 1.75 $\pm$ 0.34 & 1.19 $\pm$ 0.10 &
0.05 $\pm$ 0.14 & 16.4 $\pm$ 3.6 & $-3.47$ $\pm$ 0.35 & 1043.95 & 25.40 & 0.27 \\
C & I & 6 & $-7.33$ $\pm$ 0.18 & 1.33 $\pm$ 0.41 & 1.49 $\pm$ 0.18 & 1.74
$\pm$ 0.50 & 17.6 $\pm$ 3.67 & $-3.28$ $\pm$ 0.35 & 1038.68 & 28.05 & 1.0 \\
D & I & 5 & $-7.43$ $\pm$ 0.17 & 1.74 $\pm$ 0.39 & 1.40 $\pm$ 0.18 & ---
& 16.4 $\pm$ 3.6 & $-3.46$ $\pm$ 0.35 & 1045.34 & 22.77 & 0.5 \\
E & I & 6 & $-7.50$ $\pm$ 0.19 & 1.19 $\pm$ 0.40 & 1.43 $\pm$ 0.18 & 3.38
$\pm$ 0.15 & 16.1 $\pm$ 3.7 & $-2.99$ $\pm$ 0.29 & 1032.22 & 31.68 & 4.1 \\
F & I & 4 & $-7.52$ $\pm$ 0.17 & 1.66 $\pm$ 0.39 & --- & --- &
16.7 $\pm$ 3.6 & $-3.44$ $\pm$ 0.35 & 1051.74 & 16.37 & 0.5 \\
\hline
A & II & 6 & $-7.88$ $\pm$ 0.18 & 1.30 $\pm$ 0.38 & 2.53 $\pm$ 0.24 &
0.92 $\pm$ 0.21 & 17.4 $\pm$ 3.7 & $-3.27$ $\pm$ 0.35 & 1095.63 & 28.94 & 0.92 \\
B & II & 6 & $-7.97$ $\pm$ 0.19 & 1.75 $\pm$ 0.40 & 1.18 $\pm$ 0.16 &
0.05 $\pm$ 0.28 & 16.3 $\pm$ 3.8 & $-3.51$ $\pm$ 0.35 & 1101.29 & 23.69 & 0.75 \\
C & II & 6 & $-7.89$ $\pm$ 0.18 & 1.33 $\pm$ 0.40 & 1.37 $\pm$ 0.14 & 1.80
$\pm$ 0.50 & 17.4 $\pm$ 3.7 & $-3.29$ $\pm$ 0.35 & 1095.97 & 28.44 & 1.0 \\
D & II & 5 & $-7.97$ $\pm$ 0.17 & 1.74 $\pm$ 0.38 & 1.32 $\pm$ 0.16 &
--- & 16.3 $\pm$ 3.5 & $-3.51$ $\pm$ 0.35 & 1102.26 & 23.00 & 0.65 \\
E & II & 6 & $-8.06$ $\pm$ 0.19 & 1.26 $\pm$ 0.43 & 1.34 $\pm$ 0.15 & 3.57
$\pm$ 0.14 & 15.8 $\pm$ 3.64 & $-3.02$ $\pm$ 0.32 & 1088.53 & 31.84 & 7.5 \\
F & II & 4 & $-8.03$ $\pm$ 0.17 & 1.68 $\pm$ 0.38 & --- & --- &
16.6 $\pm$ 3.4 & $-3.49$ $\pm$ 0.34 & 1106.72 & 16.51 & 1.8 \\
\hline\hline
\end{tabular}
{\caption{\label{tab:tab-results} Best-fit parameters for the
model RLFs described in Sec.~\ref{sec:modelmethod}. $N$ is the number of free
parameters for each model, $S$ is the minimum value of eqn.~\ref{eqn:S},
Det$(\nabla\nabla S)$ is the determinant of the
{\sl Hessian} matrix evaluated at $S = S_{\rm {min}}$ and
$P_{\rm R}$ is the relative probability of the model calculated
according to eqn.~\ref{eqn:rel_prob}. 
}}
\end{center}
\end{table*}

Table~\ref{tab:tab-results} shows the relative probabilities $P_{\rm
R}$ of models A--F normalised to model C. These results demonstrate
that none of the best-fit parameterisations illustrated in
Fig.~\ref{fig:rlf_5} are unequivocally ruled out. Moreover, for all of
our models the 2D KS-test produces $P_{KS} > 0.2$ for the $L_{2.7} -
z$ plane signifying that all our models are reasonable approximations
to the data, and the 1D KS-test produces in all cases $P_{KS} \approx
0.48$ for the $\alpha$ distribution again suggesting a good working
model (see Fig.~\ref{fig:alpha_3}b). All these statements are true in
both cosmology I and cosmology II. We see that the normalisation
$\rho_{0}$ and slope of the RLF $\beta$ are consistent between the
models as expected if all the fits are reasonable. These values are
also in good quantitative agreement with those determined by DP90:
their Fig. 11 suggests $\beta \approx 1.3$, and integrating
eqn.~\ref{eqn:alpha_dist} over $\alpha$ produces values of $\rho$ at a
given $z$ within a factor of two of those derived by DP90.

Considering models A and B, we see that any marginal preference for A
largely disappears on changing from cosmology I to cosmology II.
Model A, which has a Gaussian distribution for its redshift evolution,
probably gives a false indication of the steepness of any decline in
$\rho$ at high-redshift. This is due to the coupling of the low- and
high-redshift declines introduced by the assumed Gaussian distribution: the
lack of sources below $z \leq 1$, and the flat behaviour over $1 \leq
z \leq 4$ forces the Gaussian to have a very narrow width which is
then imposed at the high-redshift end of the function. This effect is
highlighted in model B where the high-redshift evolution is characterised
by a constant co-moving space density and is not dependent on the
width of the fitted Gaussian at low-redshift.  The fitted width of
this Gaussian is extremely narrow due to the
dramatic decline at low-redshift.

Model C removes the coupling of the low- and high-redshift behaviour
present in model A but still fits a variable form at high-redshift. 
Fig.~\ref{fig:rlf_5} shows the shallowness of the decline in this model.
Comparing model C with model D, its no cut-off counterpart, we find no
evidence to suggest that the cut-off model is a significantly better
representation of the data. The relative probabilities are 0.5 and 0.65 in
favour of the Gaussian decline in cosmology I and II respectively.

Consideration of model E in which rapid (but different) declines are
enforced at low- and high-redshift, shows a steep increase in
co-moving space density from $z = 0$ to $z \approx 1.4$, followed by a
constant co-moving space density up to $z \approx 3.5$. Beyond $z
\approx 3.5$, i.e. the highest redshift object in the sample, the data
are well fitted by a very abrupt cut-off. Indeed, this has the highest
probability of all the models considered. This is suggestive of a
decline in the co-moving space density at $z \gtsim 3.5$ and provides
the only evidence from our simple parametric models for a redshift
cut-off, with the ratio of the probabilities between the worst-fit
(model B) and the best-fit (model E) models for cosmology I of
$\sim10$. There is, however, a way in which the statistical likelihood
of cut-off models might have been systematically over-estimated,
namely the effects of curvature in the radio spectra of the objects
(Sec.~\ref{sec:imp_curv}). We investigated extending our models to include
spectral curvature, but concluded that the dataset under study is
simply too small to allow the addition of further free parameters to
our existing models.

Finally, the reasonably high probability of model F illustrates the
effect of the relatively small absolute co-moving volumes observable
for the low-redshift counterparts of very rare objects. The co-moving
volume available at low-redshift ($z < 1$) is small in comparison to
that available in the redshift range $1 < z < 5$ which therefore
dominates the total volume available in both cosmologies. For $\beta \sim 1.5$, even with the decrease
in the available high-redshift volume caused by the flux-density limit
(Fig.~\ref{fig:perc_2}), we still expect to find a small fraction of
the population at low-redshift even if the space density is constant with
redshift. We will return to this point in Sec.~\ref{sec:diff_SH96DP90} in the context of
the results of SH96.

To illustrate the crucial effect of shifting the mean spectral index
we re-fitted models C (a cut-off model) and D (a no cut-off model)
from Sec.~\ref{sec:modelmethod} assuming a single radio spectral index for
the population: we calculated probabilities for $\alpha = 0.0$, the
median spectral index of the whole PHJFS sample and the value assumed
by SH96, and for $\alpha = 0.2$, the mean $\alpha$ for the most
luminous sources  In the former case the ratio of probabilities were
$\sim 50$ in favour of a cut-off model, whereas this ratio was $\sim
1$ when the true mean $\alpha$ was used. This suggests that a
parameterisation of the distribution in spectral index is essential to
any modelling of the RLF.

To summarise the results of our modelling: the only truly robust
feature of the evolutionary behaviour of the most luminous
flat-spectrum quasars is a rough constancy in $\rho$, between $1 \leq
z \leq 3.5$. At low-redshift the co-moving volume available is too small in absolute terms to be able to
make definitive comments about the space density of rare objects like
the most luminous flat-spectrum sources. At higher redshifts, the flux-density limit eats into the observable volume (Fig.~\ref{fig:perc_2})
so that again there is insufficient available volume on our light cone
to discriminate unequivocally between models with constant $\rho$ and
those with arbitrarily sharp high-redshift cut-offs.

\section{$V/V_{\mathrm {max}}$ methods}\label{sec:VVmax}
A different method of determining the extent of any evolution in the
co-moving space density of radio sources is the $V/V_{\mathrm {max}}$
statistic (Schmidt 1968; Rowan-Robinson, 1968) in which the co-moving
volume enclosed by a source is divided by the co-moving volume
available to that source given the flux-density limit of the sample and the
spectral properties of the source. To dissociate the $V/V_{\rm max}$
statistic at low-redshift from the high-redshift value we use the
banded version of the test (Avni \& Bahcall 1980; Avni \& Schiller
1983) in which these low- and high-redshift effects are
disentangled. If we define $V_{e}$ as the volume enclosed by a source
at the redshift of the source, and $V_{a}$ as the volume available to
this source, given its spectral properties, in a flux-density limited sample
then this banded version is given by
\begin{equation}
\left < \frac{V_{\mathrm e}}{V_{\mathrm a}} \right > \rightarrow
\left < \frac{V_{\mathrm e} - V_{\circ}}{V_{\mathrm a} - V_{\circ}}
\right > ,
\end{equation}
where $V_{\circ}$ is the volume enclosed at $z_{\circ}$.

\begin{figure}[!ht]
{\hbox to 0.45\textwidth{ \null\null \epsfxsize=0.45\textwidth
\epsfbox{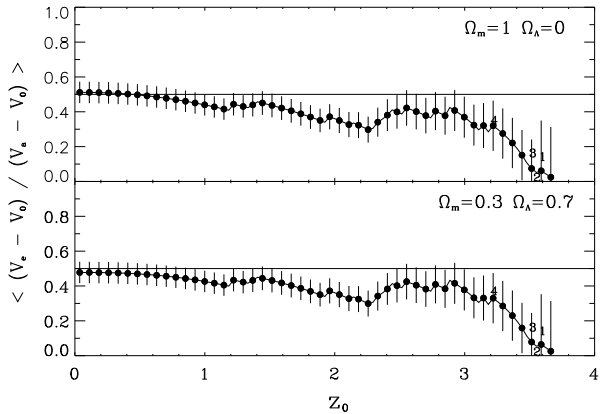} }}
{\caption{\label{fig:VVmaxbet0} Banded $<V/V_{\mathrm {max}}>$ test
for cosmology I (top) and II (bottom) with the effects of spectral curvature
explicitly considered. The vertical lines depict $1 \sigma = 1/ \surd (12N)$
error bars, where $N$ is the number of sources with redshift $>
z_{\circ}$ (see Avni \& Bahcall, 1980), and these error bars should be reasonable unless $N$ is small;
the redshifts at which $N = $ 4, 3, 2 and 1 are marked. The horizontal
line at $< (V_{\mathrm e} - V_{\mathrm e}) / (V_{\mathrm a} -
V_{\circ}) > \; = 0.5$ corresponds to the mean value for a random
distribution of sources throughout the observed co-moving volume. }}
\end{figure}

We find from the banded $V / V_{\mathrm {max}}$ test that between $z
\sim 1$ and $z \sim 2$ the points all lie $1 - 2\sigma$ below the value
expected for a non-evolving population, dropping to $\approx
2.5\sigma$ below between $2.0\, \ltsim \,z\, \ltsim\, 2.4$, before
recovering to its previous level up to $z \sim 3.2$.  At higher
redshifts the lack of sources make it very difficult to accrue any
meaningful statistics from this test, particularly as the true error
bars are, in the small numbers regime, likely to be larger and less
symmetric than those plotted (e.g. Avni \& Bahcall 1980).

One likely systematic effect is the curvature of the radio spectra of
the radio sources (Savage \& Peterson 1983).  P85 has mentioned that
it is possible to incorporate curvature into a banded $V/V_{\mathrm
{max}}$ analysis by reducing the available volume to account for a
limiting redshift beyond which a source with a given concave spectral
shape would be classified as `steep spectrum' according to a selection
criterion based on an observed spectral index. This is clearly a
factor deserving consideration, but so too is how, more generally, this
curvature will affect the observable volume: the available volume on
our light cone will normally be decreased because the observed flux-density of any source with a fixed rest-frame luminosity drops more
rapidly with redshift if it has a concave spectrum than if it has a
straight spectrum.  Our reading of the $V/V_{\mathrm {max}}$ analyses
of P85 and DP90 papers suggests that these effects have not been fully
accounted for, presumably because of the lack of data on curvature in
the radio spectra of the sources in the samples under consideration.

Our analysis of the radio spectra of the high-luminosity PHJFS sources
(Sec.~\ref{sec:imp_curv}) has allowed us to fully incorporate curvature
effects into the banded $V/V_{\mathrm {max}}$ test. We did this for
each source as follows: we evaluated the 2.7-GHz rest-frame luminosity
using the observed flux-density and a polynomial fit to the radio
spectrum; we calculate from this two limiting redshifts, one at which
a source of the same intrinsic (rest-frame) properties would have a
flux-density at the survey limit, and one at which the observed
spectrum would become steeper than the survey selection criterion; we
then calculate the available volume using the lower of the two
different determinations of critical redshift. 
The results of this banded $V/V_{\rm max}$ test (with spectral
curvature explicitly considered) are plotted in
Fig.~\ref{fig:VVmaxbet0}. We find that the systematic shift upwards in
$V/V_{\rm max}$ due to curvature effects is extremely small (less than
the symbol size in Fig.~\ref{fig:VVmaxbet0}), so P85 and DP90 were
justified in neglecting it. There is thus, robust evidence at the
$\approx 2\sigma$ level for a significant decline from the banded
$V/V_{\rm max}$ test: at $z \sim 2.2$ the statistic
is below the critical line at the $\sim 2.5 \sigma$ level, although it
approaches this line again at $z \sim 3$. The significance of the
drop at higher redshifts is still subject to worries over small number
statistics. 

In summary, there appears to be evidence for a high-redshift decline
at the $\sim 2 \sigma$ level. This is in quantitative agreement with
our likelihood analysis of Sec.~\ref{sec:modelling} where we found a ratio
of 10:1 in favour of model E with a cut-off at $z \sim 3.5$, over
models without a cut-off: a Gaussian probability distribution falls to
$\sim 1 / 10$ of its peak value roughly 2$\sigma$ away from the
location of the peak. We also find that spectral curvature has a very
small but systematic effect on the $V/V_{\rm max}$ statistic which
moves it towards the level corresponding to a randomly distributed
sample; as we will discuss in Sec.~\ref{sec:top-decade}, this effect may
become important when probing out to $z > 5$.

\section{Constraining the uncertainties of any high-redshift decline}\label{sec:cons_uncert}
To further quantify the significance of any high-redshift 
cut-off we have modified the parametric modelling of Sec.~\ref{sec:modelmethod}
and used additional variants of the $V/V_{\mathrm {max}}$ test
used in Sec.~\ref{sec:VVmax}.

The modified likelihood analysis assumes a very simple high-redshift
distribution of the form
\begin{equation}\label{eqn:eta}
\rho_{z}(z) \propto \left (\frac{1+z}{1+z_{\rm {peak}}}\right )^{\eta},
\end{equation}
in which we fix the peak redshift $z_{\mathrm {peak}} = 2.5$, and use 
the fitted values given in Table.~\ref{tab:tab-results} for the luminosity
function $\beta$, the spectral index distribution $\gamma$ and
$\epsilon$. We find the probability distribution function (pdf) for $\eta$ 
by integrating over all values of the normalisation parameter 
$\rho_{\circ}$, i.e.
\begin{eqnarray}
P(\eta \,| \,{\rm data}) = \int P({\rm data}\, |\, \log_{10} \rho_{\circ},\, \eta
) \times \nonumber \\ P(\log_{10} \rho_{\circ}) \,{\rm d}(\log_{10}
\rho_{\circ}),
\end{eqnarray}
where $P({\rm data}\, |\, \log_{10} \rho_{\circ},\, \eta )$ is proportional
to the likelihood function, ${\mathcal L} \propto \exp^{-S/2}$, and we
have assumed a uniform prior for $\eta$. $S$ is found by
minimising eqn.~\ref{eqn:S} with a luminosity function $\rho$ of the form,
\begin{equation}
\rho = \rho_{\circ} \times \rho_{L}(L_{2.7}) \times \rho_{\alpha}(\alpha) \times \left (\frac{1+z}{1+z_{\mathrm peak}}\right )^{\eta}.
\end{equation}
Note that this is an approximation because we are 
fixing the majority of the
parameters and only integrating over $\rho_{\circ}$. However this
is reasonable in this case as the fixed parameters will not be as
strongly correlated with $\eta$ as $\rho_{\circ}$.

\begin{figure}[!ht]
{\hbox to 0.45\textwidth{ \null\null \epsfxsize=0.45\textwidth
\epsfbox{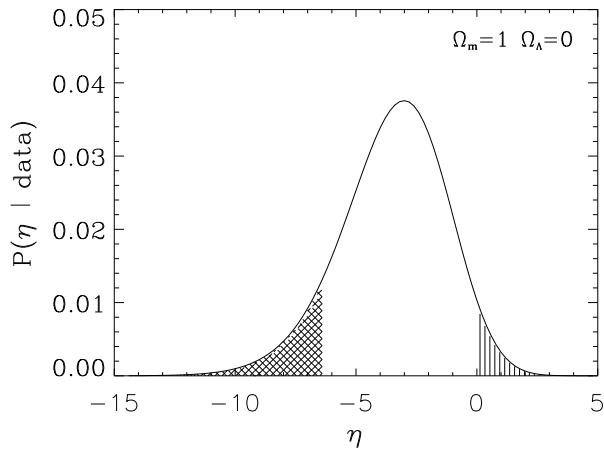}
}} {\caption{\label{fig:gammaxlike} The probability distribution for
$\eta$ from the maximum likelihood method of Sec.~\ref{sec:cons_uncert},
adopting cosmology I. The meshed region corresponds to the region
where $\eta$ is steeper than or equal to the abrupt decline preferred
by SH96 and SH98; the region shaded with vertical lines corresponds to
the values of $\eta$ where there is either a constant or increasing
co-moving space density.}}
\end{figure}

Fig.~\ref{fig:gammaxlike} shows the pdf determined for $\gamma$ from
this maximum likelihood method. The area under the curve at $\eta \ge
0$ indicates that there is a detection of a decline above $z \geq 2.5$
with $95\%$ confidence. Thus, although one cannot unequivocally rule
out a constant co-moving space density using this method, it is in
quantitative agreement with the methods of Sec.~\ref{sec:modelling} and
Sec.~\ref{sec:VVmax}, and highly suggestive of some high-redshift decline.
We also estimate the probability of the very abrupt decline ($\eta
\approx -6.5$; corresponding to a $\sim 1.5$ dex decline between
$z\sim 2.5$ and $z\sim 5$) preferred by SH98. We find from the area
under the curve that this is ruled out with $92\%$ confidence.  Thus,
although we cannot rule out a decline as abrupt as that envisaged by
SH96 and SH98, we find it is disfavoured.  The peak of the
distribution agrees with a shallow decline reminiscent of the results
of DP90. However, the presence of spectral curvature has been
neglected by this method which as we now demonstrate systematically
shifts the distribution to less negative values of $\eta$.

\begin{table}[!ht]
\small
\begin{center}
\begin{tabular}{c|c|c|c|c}
\hline\hline 
\mc{1}{c|}{$\eta$} & \mc{1}{c|}{Weighted $V/V_{\mathrm {max}}$} &
\mc{1}{c|}{$P_{\mathrm {KS}}$} & \mc{1}{c|}{Cosmology} & \mc{1}{c|}{Curvature} \\
\hline\hline
-6.50 & 0.64 $\pm$ 0.10 & 0.41 & I & No \\
-3.66 & 0.50 $\pm$ 0.10 & 0.56 & I & No \\
0.00  & 0.24 $\pm$ 0.10 & 0.15 & I & No \\
\hline
-6.50 & 0.67 $\pm$ 0.10 & 0.31 & I & Yes \\
-2.77 & 0.50 $\pm$ 0.10 & 0.57 & I & Yes \\
0.00  & 0.34 $\pm$ 0.10 & 0.44 & I & Yes \\
\hline\hline
-6.50 & 0.64 $\pm$ 0.10 & 0.41 & II & No \\
-3.64 & 0.50 $\pm$ 0.10 & 0.56 & II & No \\
0.00  & 0.25 $\pm$ 0.10 & 0.17 & II & No \\
\hline
-6.50 & 0.66 $\pm$ 0.10 & 0.32 & II & Yes \\
-2.73 & 0.50 $\pm$ 0.10 & 0.57 & II & Yes \\
0.00  & 0.35 $\pm$ 0.10 & 0.45 & II & Yes \\
\hline\hline
\end{tabular}
{\caption[junk]{\label{tab:VVmaxtable} Results of the weighted
$V/V_{\mathrm {max}}$ tests described in Sec.~\ref{sec:VVmax}. \\
$\eta_{\mathrm {max}} = -3.66_{-1.83}^{+1.47}$ with no curvature for
cosmology I.\\
$\eta_{\mathrm {max}} = -2.77_{-2.04}^{+1.74}$ with curvature for
cosmology I. \\
$\eta_{\mathrm {max}} = -3.64_{-1.86}^{+1.51}$ with no curvature for
cosmology II. \\
$\eta_{\mathrm {max}} = -2.73_{-2.08}^{+1.79}$ with curvature for
cosmology II. \\ 
The upper and lower limits on $\eta_{\mathrm {max}}$ have been
calculated according to the prescription of Avni \& Bahcall (1980), in
which values of $\eta$ are evaluated $1\sigma$ away from the mean $V/V_{\mathrm
{max}}$, again assuming Gaussian errors with a standard deviation $\sigma
= (\sqrt{12N})^{-1}$.}}
\end{center}
\end{table}

To quantify the effect of spectral curvature we again turn to a variant of the 
$V/V_{\mathrm {max}}$ test discussed by Avni \& Bahcall (1980).
We use a weighted $V/V_{\mathrm {max}}$ statistic, i.e. 
\begin{equation}
<A> = \frac{\int_{V_{\circ}}^{V_{\mathrm e}} \rho_{z}
(z) {\rm d}V}{\int_{V_{\circ}}^{V_{\mathrm a}} \rho_{z} (z) {\rm d}V},
\end{equation}
where $\rho_{z} (z)$ is again given by eqn.~\ref{eqn:eta}, $V_{\circ}$ is
the volume enclosed at the peak redshift (again set at $z=2.5$),
$V_{\mathrm e}$ and $V_{\mathrm a}$ are the volume enclosed by the
source and the volume available to the source respectively. Following
Avni \& Bahcall (1980), we calculate this statistic as a function of
$\eta$ and determine a best-fit value at the point $<A> = 0.5$. The
results of this investigation, with and without taking spectral
curvature into account, are presented in Table~\ref{tab:VVmaxtable}.

It is obvious from Table~\ref{tab:VVmaxtable} that accounting for
spectral curvature changes the best-fit slightly,
favouring a more gradual drop in the value of $\rho$ at high
redshifts, but does not remove the evidence for a decline. We
conclude that a gentle decline in $\rho$, amounting to a factor $\sim
4$ from $z \sim 2.5$ to $z \sim 5$, is favoured by the data, with both
constant space density models and abrupt cut-off models ruled out at
roughly the $2 \sigma$ level.
\begin{figure*}
{\vbox to 0.85\textheight{\null\null
{\hbox to \textwidth{ \null\null 
\epsfxsize=0.6\textwidth
\epsfbox{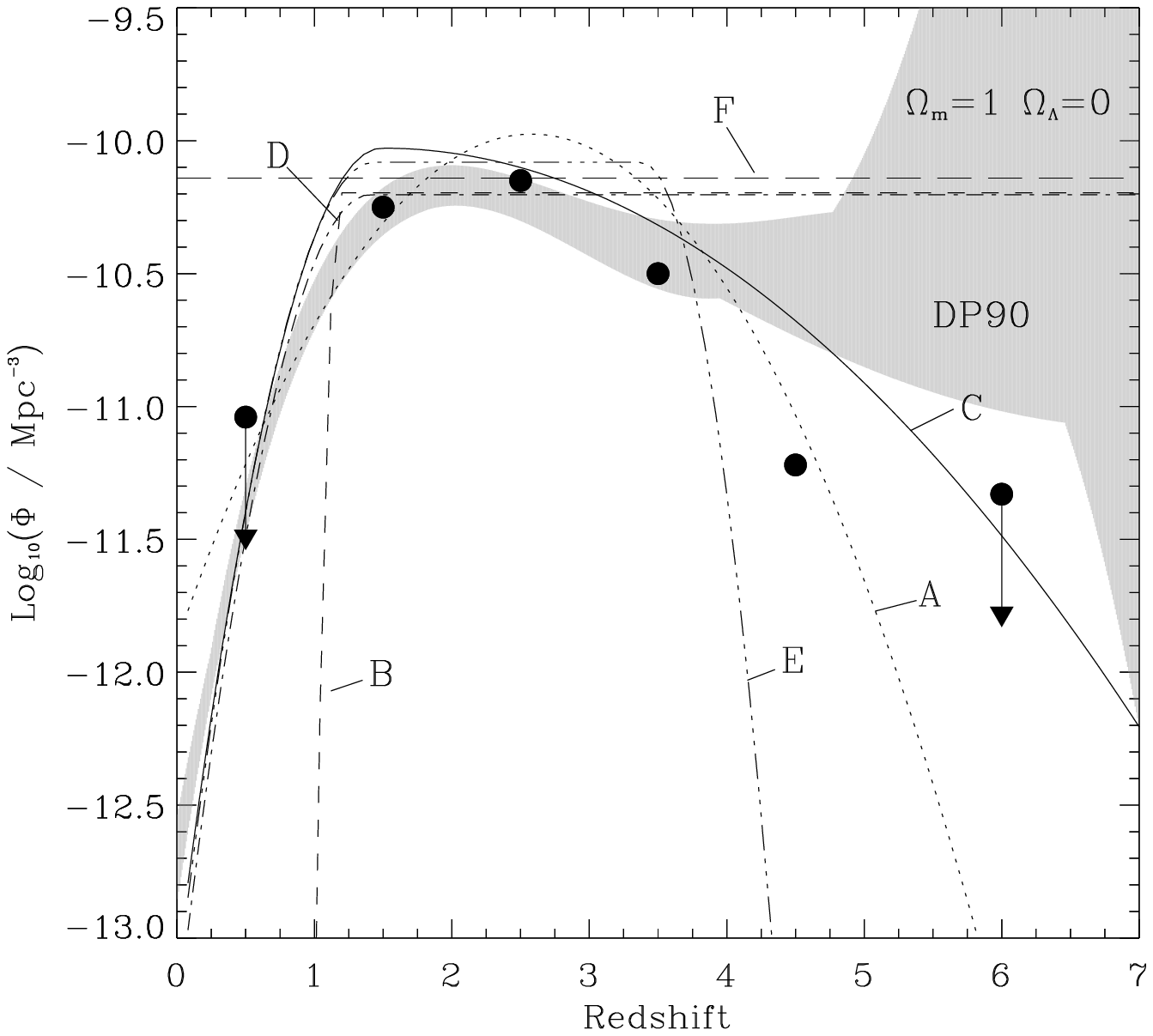}
}}
\vspace{0.5cm} {\hbox to \textwidth{ \null\null
\epsfxsize=0.6\textwidth
\epsfbox{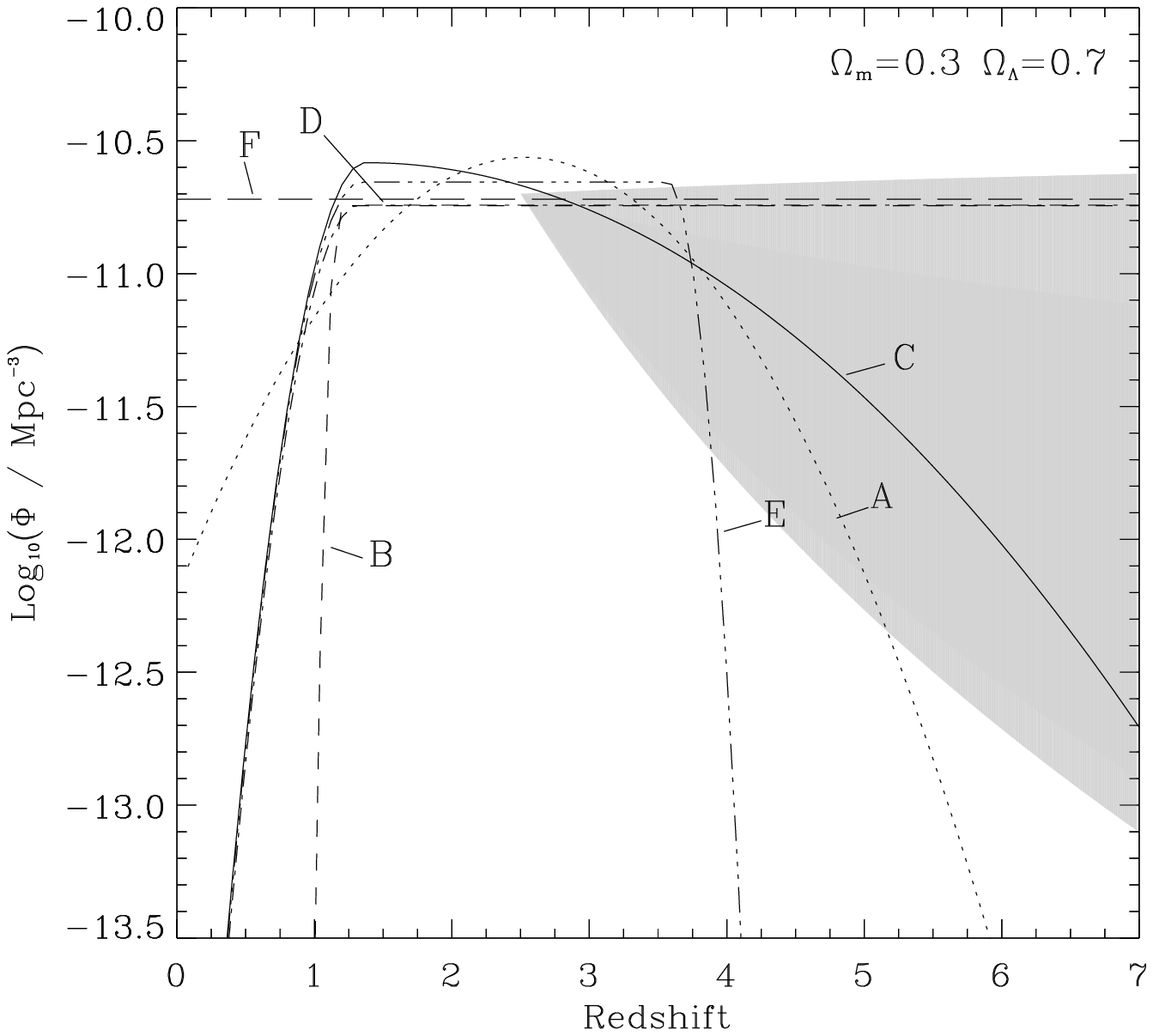} }} }}
{\caption{\label{fig:rlf_5} (a) top: The co-moving space density
$\Phi$ ($\int\int\int\rho\,{\rm d}(\log_{10}L_{2.7}){\rm d}z{\rm
d}\alpha$\,) for the six model radio luminosity functions described in
Sec.~\ref{sec:modelmethod} for cosmology I. The range of free-form RLF
models considered by DP90 are also shown by the shaded region for
cosmology I. The filled circles show the binned points of SH98 which
were calculated assuming $\alpha = 0$ and using a binned method; note
the points at $z = 0.5$ and $z = 6.5$ are upper limits which
correspond to finding one object at these redshifts. (b) bottom: The
co-moving space density for the six model RLFs described in
Sec.~\ref{sec:modelmethod} for cosmology II. The shaded region corresponds
to the 90 per cent confidence region of the high-redshift space
density of quasars using the $V/V_{\mathrm {max}}$ method with
curvature included as discussed in Sec.~\ref{sec:cons_uncert}. }}
\end{figure*}
\section{Discussion}\label{sec:discussion}
\subsection{Understanding the differences between the DP90 and SH96 results}\label{sec:diff_SH96DP90}
In Fig.~\ref{fig:rlf_5}a we have compared our set of parametric models
(Sec.~\ref{sec:modelling}) with the results of the free-form analysis
of DP90 (for the top decade of their flat-spectrum RLF), and the
binned evaluation of SH96. Our models A and E look very like the SH96
points, and our models B and D look very like the DP90
results. Although our analysis of Secs.~\ref{sec:modelling},
~\ref{sec:VVmax} and ~\ref{sec:cons_uncert} suggest that it is
currently not possible to discriminate unequivocally between these
very different high-redshift behaviours, it does lead us to be
confident (at the $\sim 95$ \% level) that the DP90 results are the
more reliable.  The abruptness of the decline suggested by SH96 and
SH98 seems likely to be the result of three factors: (i) use of
$\alpha = 0$ as a representative spectral index, whereas the most
luminous flat-spectrum sources are significantly steeper; (ii) use of
a binned method which does not fully account for the steepness of the
RLF across the bin which is particularly important at high-redshift
where the flux-density limit cuts into the observable volume; and
(iii) no corrections for radio spectral curvature.  These systematic
effects have been discussed in detail in Secs.~\ref{sec:imp_spix}
and~\ref{sec:imp_curv}, and since they all work in the same direction it
seems clear that they can combine to produce values of $\rho$ which
are systematically biased to low values, possibly by a large factor.

The band of DP90 free-form models diverges rapidly at high-redshifts,
but seem at face-value to rule out a very rapid decline (declining by
at most a factor $\sim 3$ between $z = 2.5$ and $z = 5$). Our analysis
(Secs.~\ref{sec:VVmax} and~\ref{sec:cons_uncert}) has attempted to assign
probabilities to the strength of this decline which, considering a
90\% confidence interval, allow a fairly broad range of declines at
high-redshift (see Fig.~\ref{fig:rlf_5}b). Since the direct
constraints on the top decade of the RLF are
very similar between this work and DP90 (see
e.g. Table~\ref{tab:samples} and Fig. 9 of DP90), this is simply an
artifact of differences between our method and the free-form method of
DP90.

All likelihood methods applicable to sparsely populated datasets (like
those considered here) require some assumption of smoothness, be it a
simple functional form (Sec.~\ref{sec:modelling}) or a series expansion
(DP90). This means that there will be cross-talk between the
$\log_{10} (L_{2.7})$-$z$ range of interest and other parts of the
$\log_{10} (L_{2.7})$-$z$ plane.  The worry is that if, for example, the top
decade of the RLF was dominated by a physically-different phenomenon
to the rest of the RLF; then cross-talk could introduce abhorrent
features into the RLF under investigation. In
Secs.~\ref{sec:linking_steep} and~\ref{sec:cosmo_context} we will show that
this may be a real concern.

However, cross-talk can be beneficial, as the following discussion
illustrates.  Consider the low-redshift part of Fig.~\ref{fig:rlf_5}.
Both the binned analysis of SH96 and the DP90 modelling suggest very
firm evidence for a decline at low-redshifts, whereas the reasonably
high relative likelihood of our model F (Sec.~\ref{sec:modelresults})
seems to require no such behaviour. The difference between the SH96
point and our model F can no longer be ascribed to the systematic
effects discussed in Secs.~\ref{sec:imp_spix} and~\ref{sec:imp_curv} since, as
illustrated by Fig.~\ref{fig:perc_2}, all the most radio-luminous
flat-spectrum sources on our light cone are detectable in the area
surveyed. The difference is largely due to small number
statistics. The ratio of volumes between the SH96 redshift bins
centred at 0.5 and 1.5 is about $1 / 3$ (cosmology II) \footnotemark,
so from Table~\ref{tab:tab-results} and the assumption of a fairly
constant $\rho$, $\sim 2$ sources would be predicted in the
low-redshift bin whereas zero are seen. This has a Poisson probability
($\sim \!e^{-2}=0.13$), and its true significance must also allow for
Poisson noise in the upper redshift bin (which accounts for the drop
in the normalisation of model F with respect to the other models), so
that as a detection of a `low-redshift' cut-off it is not particularly
significant.  However, the tight constraints on the DP90 models give
firm evidence for a significant low-redshift decline, and although
this is an artifact of cross-talk, it is almost certainly true. The
reason for this is as follows: if we were to gradually lower the
critical luminosity (arbitrarily) defining our `most radio luminous'
sub-set, then the number of such sources at low-redshift would
increase rapidly by virtue of the steep radio luminosity function; in
a repeated binned analysis, the effects of Poisson noise could be
marginalised, illustrating a beneficial effect of the cross-talk
inherent in the DP90 method. \footnotetext{This is the one calculation
in this paper where the difference between cosmologies I and II has an
important effect, so we prefer the one currently favoured by
observations (e.g. Perlmutter et al. 1999). The {\em relative}
co-moving volumes between two redshift bins are relatively large if
the bins are at low- and high-redshift, but small if both bins are at
high-redshift (which is always the case for the most luminous PHJFS
sources considered in this paper).}  

The crucial point is that similar arguments -- namely solving the
problem of small number statistics by extending the study to include
lower luminosity objects -- are not applicable in anything like the
same straightforward manner to any high-redshift cut-off. The
situation is greatly complicated by the truncations enforced on the
observable volume by the existence of flux-density selection limits
(see Secs.~\ref{sec:imp_spix} and~\ref{sec:imp_curv}, particularly
Fig.~\ref{fig:perc_2}). Because of these truncations, the search for
low luminosity objects at high-redshifts requires the use of fainter
samples. The work of DP90 utilised $L_{2.7} - z$ data from a fainter
sample (see Table~\ref{tab:samples}) which did allow some sensitivity
to less radio luminous objects at high-redshift, but because of the
small sky area of this sample, the total number of $z >2$
flat-spectrum sources in their study is similar to that in the PHJFS
sub-set studied here. Given this fact, and given that a significant
fraction of sources are common to both studies
(Sec.~\ref{sec:PHJFSsample}) it is no surprise that, for example, their
$V/V_{\mathrm {max}}$ analysis yields strikingly similar results
(compare Fig.~\ref{fig:VVmaxbet0} with Fig. 12 of DP90). We note also
the worrying incompleteness of the faintest DP90 sample near the
flux-density limit (DP90 Appendix A): this is a particular problem for
interpreting low values of $V/V_{\mathrm {max}}$ as evidence for a
high-redshift cut-off, since objects with high values of $V/V_{\mathrm
{max}}$ will always lie near the flux-density limit of a survey, and if these
sources are in the top-decade of the RLF they must be at
high-redshift.

To conclude this section we contend that the evidence against models
with constant space density at high-redshift, or indeed a cut-off as
abrupt as that envisaged by SH96 and SH98, is not yet compelling.  The
evidence for any cut-off at high-redshift for flat-spectrum sources is
at present significant only at the $\sim 2 \sigma$ level, and
therefore in our view still tentative.  However, the best bet, as
quantified in Fig.~\ref{fig:rlf_5}b and Table~\ref{tab:VVmaxtable} and
as previously suggested by P85 and DP90, is that such a cut-off does
exist, and amounts to a gradual decline in $\rho$ by a factor $\sim 4$
between redshifts $z \sim 2.5$ and $z \sim 5$.

\subsection{Linking the most luminous flat-spectrum quasars with the
steep-spectrum population}\label{sec:linking_steep}
The most common physical
cause of a flat radio spectrum at $\sim \!1$\,GHz frequencies is
synchrotron self-absorption implying that the emission comes from
compact regions. This means that there are at least two distinct ways
in which the steep- and the flat-spectrum populations could be linked.
The first way concerns the early phases of radio source evolution in
which the emission comes from regions so compact that a Giga-Hertz
Peaked radio Spectrum (GPS) inevitably results. Current theories
(see review by O'Dea 1998) suggest that all FRII (Fanaroff-Riley class
II; Fanaroff \& Riley 1973) radio sources (before developing
their large-scale steep-spectrum lobes) pass through a GPS phase,
and move through a Compact Symmetric Object (CSO) phase as the source
further expands and its turnover frequency drops. The second way concerns
sources favourably oriented such that the emission from the base of
one of their jets is beamed along the line-of-sight, producing
Doppler-Boosted (DB), or `core-dominated', sources (e.g. Blandford \&
Rees 1978). We will consider these two possibilities in turn.

To what part of the steep-spectrum population would GPS sources with
$\log_{10} (L_{2.7}) > 27$ correspond? There are two strong lines of
argument that for a given source $L_{2.7}$ must decline steadily with
time. First assuming that the environmental density declines with
radius, any reasonable model for source expansion (e.g. Begelman 1996;
Kaiser \& Alexander 1997; Blundell, Rawlings \& Willott 1999; Blundell
\& Rawlings 1999) predicts this must be the case. Secondly, if the
luminosity was constant or increased with time then the known GPS
sources would dramatically over-produce FRIIs.  A synthesis of these
ideas (O'Dea \& Baum 1997) suggests that $L_{2.7} \propto D^{-0.5}$,
so that in growing out of the GPS phase (at a size $\sim 0.3$ kpc)
into a FRII (of size $\sim 300$ kpc), $L_{2.7}$ declines by a factor
$\sim 10^{1.5}$, and since the source will now be optically thin at
much lower frequencies we can extrapolate this value (using $\alpha =
0.8$) to obtain $L_{151} = 26.5$. This is well above the FRI/FRII
break, so the most-luminous GPS sources seem certain to become FRIIs,
and they lie at or just above the $L_{151}$ boundary at which the
quasar fraction of FRIIs drops precipitously (e.g. Willott et
al. 2000a).  Measurements of the RLF suggest that sources of this
$L_{151}$ have a space density $\log_{10} \rho \sim -6.5$ at $z \sim
2.5$ (Willott et al. 2000b; DP90), roughly 3 dex higher than the value
of $\rho$ inferred for the $\log_{10} (L_{2.7}) > 27$ flat-spectrum
population.  We therefore have no difficulty linking this population
with the $\log_{10} (L_{151}) \ltsim\ 26.5$ FRIIs providing the GPS phase
persists for a time which is $\sim 10^3$-times shorter than the FRII
phase. This seems plausible since estimates of the kinematical ages of
FRII radio sources, based on lobe asymmetries and light travel-time
effects, gives a value $\sim 10^{7.5}$ yr for a 500-kpc FRII
(e.g. Scheuer 1995), whereas a direct measurement of the expansion
rate of a GPS source by Owsianik \& Conway (1998) suggests a
kinematical age in the range $10^{3-4}$ yr.

To what part of the steep-spectrum population would DB sources with
$\log_{10}(L_{2.7}) > 27$ correspond? An attempt to unify the flat-
and steep-populations via Doppler boosting has recently been made by
Jackson \& Wall (1999).  They find that to produce flat-spectrum
sources from a parent FRII requires that the jets have a bulk Lorentz
factor $\gamma \approx 8.5$, and requires that the jets are aligned
within a critical angle $\theta_{\rm crit} \approx 7^{\circ}$ of the
line-of-sight. These values set a characteristic Doppler
factor\footnote{Doppler factor $\Gamma = \gamma^{-1} (1 - \beta \cos
\theta)^{-1}$, where $\beta$ is the speed (in units of $c$) of the
bulk motion and $\theta$ is the angle of the motion with respect to
the line-of-sight.} $\Gamma\, \sim\, 10$ for the luminous
flat-spectrum sources, and hence predict a boosting of the core flux
by a factor $\Gamma^{2}|_{\theta = \theta_{\rm crit}} /
\Gamma^{2}|_{\theta = \theta_{\rm trans}} \sim 10^{3}$ in a comparison
between flat spectrum objects and their counterparts aligned at the
transition angle $\theta_{\rm trans}$ where, according to unified
schemes, the quasar nucleus just becomes optically visible (we take
$\theta_{\rm trans} \approx 53^{\circ}$ from Willott et al. 2000a).
Measured core-to-lobe ratios $R$ at high radio frequencies are $\sim
10^{-1.5}$ for objects marking the division between quasars and radio
galaxies, i.e. those aligned at $\theta \approx \theta_{\rm trans}$
(see Fig.\ 8 of Jackson \& Wall 1999). For DB sources the core is
likely to be boosted to a flux-density much ($\sim 10^{1.5}$) higher
than the (presumably roughly isotropic) high-frequency flux-density of
the lobe.  This means that a $\log_{10}(L_{2.7}) > 27$ source will have a
lobe with $\log_{10} (L_{2.7})\, \gtsim\, 25.5$, or (taking $\alpha=0.8$)
$\log_{10} (L_{151})\, \gtsim\, 26.5$.  In other words, we expect both the
GPS and DB populations to be drawn from similar parts of the RLF of
the FRII population, one dex above the FRI/FRII divide, and about at
the point where the observed quasar fraction switches from $\approx
\!0.1$ to $\approx \!0.4$ (Willott et al. 2000a). As for the GPS
objects, the DB objects will be drawn from a population with a space
density $\log_{10} \rho\, \sim\, -6.5$ at $z \sim 2.5$ (Willott et
al. 2000, DP90), but this time it is a small beaming angle, rather
than a small source age which lies behind the lower derived $\rho$ for
the flat-spectrum population.  Taking $\theta = 7^{\circ}$ means that
only $\sim 1$ in $\sim 200$ sources is expected to be favourably
oriented, but this still slightly over-produces the observed
flat-spectrum population at $\log_{10}(L_{2.7}) > 27$. There are enough
uncertainties in this argument that this is not a serious problem: for
example, adopting a slightly lower characteristic Doppler factor would
allow us to associate the DB objects with slightly more radio luminous
FRII parents, say $\log_{10} (L_{151})\, \gtsim\, 27$, and thus resolve the
problem.  An important corollary of these ideas will be discussed in
Sec.~\ref{sec:less_lum}.

With the possibility of a mixed GPS and DB population in mind, the
obvious next step was to turn our attention to the radio properties of
the $\log_{10} (L_{2.7}) > 27$ PHJFS sample. In
Fig.~\ref{fig:radspec_6} we have shown radio spectra of each object,
together with rough spectral classifications.  Simple inspection shows
that in several cases where the data are adequate, the spectra are
characteristic of GPS or CSS sources (four of these objects are in the
O'Dea review on GPS sources and 2 others are present in the GPS sample
of de Vries, Barthel \& O'Dea 1997). Few of the spectra bear much
resemblance to the compilation of DB objects of Gear et al.(1994)
which typically peak around 10\,GHz and decline at higher
frequencies. The nature of the minority of objects with straight or
concave spectra is unclear, although they are reminiscent of the
Core-Jet Sources (CJS) mentioned by Willott et al. (1998), the
archetype being 3C286, which, having jet structures on {\em both}
sides of the nucleus, are believed to be dominated by a component
which lacks strong Doppler boosting.

\begin{figure}[!ht]
{\hbox to 0.45\textwidth{ \null\null \epsfxsize=0.45\textwidth
\epsfbox{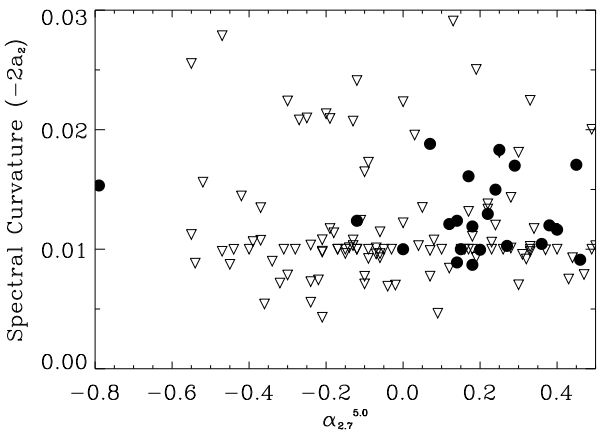} }}
{\caption{\label{fig:betaalp} The distribution in the curvature
($-2a_{2}$) of radio spectra as a function of $\alpha_{2.7}^{5.0}$ for the
most luminous PHJFS sources (filled circles) and sources from the next lower
decade in radio (2.7\,GHz) luminosity (triangles). }}
\end{figure}

To pursue this qualitative result, we plot in Fig.~\ref{fig:betaalp}
the curvature ($-2a_{2}$, where $a_{2}$ is defined in
Sec.~\ref{sec:imp_curv}) of the radio spectrum against observed radio
spectral index for the most luminous PHJFS sub-set, and the PHJFS
sources from the next lower dex in radio luminosity. The shift in
observed spectral index for the most radio-luminous sources
(Sec.~\ref{sec:imp_spix}) is clearly shown in this plot, but we also see
hints that the spectral curvatures are significantly different for the
most luminous subset (naive application of the 1-D KS test yields
$P_{\rm KS} = 0.02$ for the distributions in spectral curvature of the
two samples, with median curvature $-2a_{2} = +0.34$ in the top decade
and $-2a_{2} = -0.14$ in the next lower decade). It is not possible to
make more definitive statements from this plot because the radio data
used to construct the radio spectra are rather inhomogeneous (we have
compiled them using data from the Nasa Extra-galactic Database),
particularly in terms of their frequency coverage: some of the objects
may be characterised as having straight spectra ($-2a_{2} \sim 0$), or
highly curved spectra, merely because of the lack of many frequency
points in their radio spectra and/or time variability. However, it
does seems clear that there are a significant number of GPS sources in
the top decade of the flat-spectrum RLF, and more arguably they seem
to form a lower fraction of the population at lower radio
luminosities. Accepting these arguments allows us to tie up two loose
ends. First, the shift towards steeper radio spectra in this decade
(Fig.~\ref{fig:alpha_3}) can now be understood as the emergence of a
population of GPS source with turnover frequencies in the rest-frame
GHz range, and hence characteristically steeper spectra at an observed
frequency of 2.7\,GHz.  Second, the worries concerning cross-talk
between lower parts of the RLF, and its top decade
(Sec.~\ref{sec:top-decade}) inherent in any analysis assuming a smooth
continuous RLF appear to be worthy of further investigation. As we
will show in Sec.~\ref{sec:less_lum} a significant ratio of GPS-like
sources to DB sources is likely to be restricted to a fairly narrow
range of 2.7\,GHz luminosity so extrapolations of any properties,
including space density, must be pursued with extreme care in the
overlap region between the two physically distinct populations.
\section{Flat-spectrum quasar evolution in a cosmological context}\label{sec:cosmo_context} 
In Sec.~\ref{sec:cons_uncert} we discussed that the evidence of any
decline in the $\rho$ of flat-spectrum quasars at high-redshifts is
rather gentle compared to the evolutions preferred by the analyses of
SH96 and SH98.  This is entirely consistent with constraints on the
high-redshift evolution in $\rho$ for the steep-spectrum population to
which, as argued in Sec.~\ref{sec:linking_steep}, this flat-spectrum
population is ultimately linked.  Willott et al. (1998) have determined
that quasars with $\log_{10} (L_{151})\, \gtsim \, 27$ rise in space
density by $\sim 2$ dex between $z \sim 0$ and $z \sim 2$, but then
stay at roughly constant $\rho$ out to $z \sim 3$, beyond this
redshift the evolution in $\rho$ remains poorly constrained. This
mirrors the behaviour of the entire FRII population at these
luminosities (DP90, Willott et al. 2000b) which is unsurprising given
that the quasar fraction shows no dependence on radio luminosity at
$\log_{10} (L_{151}) > 26.5$ (Willott et al. 2000a). Efforts targeted
at mapping the high-redshift cut-off of the steep-spectrum population
have yet to find any evidence of a high-redshift decline in the $\rho$
of the steep-spectrum FRII population (Rawlings et al. 1998; Jarvis et
al 1999; Willott et al. 2000b; Jarvis et al. 2000), but are consistent
with a gentle decline in $\rho$ (as favoured for the flat-spectrum
population in this paper and in DP90).

How does this situation compare to the cosmic evolution of
optically-selected quasars? It has now been established that the
optical luminosity function (OLF) is steep ($\beta \sim 1.5$) with an
increase in $\rho$ by $\approx \!2$ dex out to $z \sim 2.2$
(e.g. Goldshmidt \& Miller 1998). At higher redshifts, despite their
different selection techniques, several surveys have now yielded
fairly consistent results favouring a significant decline at high
redshift: Warren, Hewett \& Osmer (1994) from a multicolour survey,
Schmidt, Schneider \& Gunn (1995) using quasars discovered by their
Ly$\alpha$ emission, Hawkins \& Veron (1996) using a variability
technique, and Kennefick, Djorgovski \& de Carvalho (1995) using a
multicolour technique have all argued for an abrupt decline in $\rho$
between $z \sim 2.5$ and $z \sim 4$. As shown on
Fig. 2 of SH96, these results are all consistent with a decline in
$\rho$ by a factor $\sim 10$ between $z \sim 2.5$ and $z \sim
5$. \footnotemark \footnotetext{ A contrary view has been put forward
by Irwin, McMahon \& Hazard (1991) who discussed evidence that $\rho$
for the optically brightest quasars is roughly constant between $z \sim 2$ and
$z \sim 4.5$.  The optical and UV spectra of quasars are far more
complex than the relatively simple radio spectra studied in this
paper, and uncertainty in $K$-corrections could lead to complications
in determining the absolute magnitude of sources needed to correctly
place quasars at various redshifts from various samples onto the steep
OLF. So, just as $K-$corrections were vital in understanding the
discrepancies between previous determinations of the flat-spectrum RLF
(Sec.~\ref{sec:cons_uncert}), this may also prove to be the case for
optically-selected quasars.  As emphasised recently by Wolf et al. (1999), different optical selection techniques for quasars may also
need to be explored.  } Comparing the redshift cut-off of
optically-selected quasars with the cut-off for flat-spectrum radio
quasars, as put forward here and in DP90, suggests that the drop in
$\rho$ is more abrupt for the optically-selected population than for
the radio-selected population, whereas SH96 and SH98 argued that the
declines are similar.  One obvious way in which $\rho$ might
drop more rapidly for optically-selected quasars than for
radio-selected quasars would be if there is an increasing chance of
dust obscuration at high-redshift, either dust in intervening systems
(e.g. Fall \& Pei 1989), or dust associated with their young host galaxies.

X-ray surveys of quasars provide another probe of evolutions in
$\rho$.  Both Boyle et al. (1994) and Page et al. (1996) successfully
fitted a pure luminosity evolution (PLE) model to the X-ray luminosity
function (XLF) with the evolution slowing down, but not necessarily
reversing above a redshift $z \sim 1.7$. The most recent results by
Miyaji et al. (2000), using just $ROSAT$ observations to avoid
cross-calibration problems, found similar behaviour, and no evidence
for any high-redshift cut-off. Our results of Sec.~\ref{sec:cons_uncert}
show that any discrepancy between the high-redshift evolution of X-ray
selected quasars and flat-spectrum radio quasars which were
highlighted by Miyaji et al. (2000) are resolvable within the
uncertainties of both samples.

In a still wider cosmological context, the redshift cut-off of active
galaxies has been linked to the cosmic evolution in the global
star-formation rate (e.g. SH98). The most recent studies of the global
high-redshift star-formation rate based on (dust corrected) optical
data (Steidel et al. 1999) and on sub-mm data (e.g. Blain et al. 1999,
Eales et al. 1999, Hughes et al. 1998) are now in reasonable accord: the
Steidel et al. study finds no significant change in the luminosity
function of Lyman-break galaxies between redshifts $z \approx 3 $ to
$z \approx 4$, and thus no evidence yet for a high-redshift decline in
global star formation. Our results of Sec.~\ref{sec:cons_uncert} remove
any significant difference between global star formation and the
evolution of flat-spectrum radio sources, as the form of the
high-redshift evolution in both cases is consistent within the
uncertainties. Differences inferred from comparing the Steidel et al. work and
optically-selected quasars are probably robust (see e.g. Cen 2000) but
arguably now should be discussed in the context of possible dust
obscuration of the optically-selected quasar population at high
redshift.

\section{Can the redshift cut-off for flat-spectrum quasars 
ever be firmly established?}\label{sec:estab_cutoff}

\subsection{The top decade of the flat-spectrum RLF}\label{sec:top-decade}

We first assess the chances of establishing beyond doubt a redshift
cut-off using just the most luminous flat-spectrum quasars.  The SH96
approach to this problem was to search for $z > 5$ quasars in their
faintest 2.7-GHz radio survey which has a flux-density limit of
$S_{2.7} > 0.25$\,Jy (see Table~\ref{tab:samples}).  They found no $z
> 5$ quasars, and in SH98 say that if there was no fall-off in the
space density of quasars at high-redshift, they would have expected to
find 15 quasars at $5 < z < 7$.

We will compare the predictions 
of models C and D from Sec.~\ref{sec:modelling}, 
the former having a fairly gradual decline at high-redshift, and the
latter having a co-moving space density which remains constant above a
peak $z$\footnote{The results of this section do not depend critically 
on the assumed cosmological model for the reasons discussed in 
Sec.~\ref{sec:diff_SH96DP90}; we use Cosmology I}. The
high-redshift behaviour of these models roughly correspond to the
$\sim 90$ per cent confidence interval derived in
Sec.~\ref{sec:cons_uncert} and plotted on Fig.~\ref{fig:rlf_5}b.

We have made Monte Carlo simulations of a survey similar to the
Parkes-based work of SH96 and SH98, i.e. a 2.7-GHz flux-density limit of
$0.25\,$Jy over a sky area of 3.8\,sr with a spectral index selection
criterion $\alpha_{2.7}^{5.0} \leq 0.4$.  These simulations show that
we would expect on average to find $0.3$ objects with $\log_{10}
(L_{2.7}) \geq 27$ in the redshift range $5 \leq z \leq 7$ for model C
and $4$ objects for model D.  The corresponding numbers of quasars
predicted in the $4 \leq z \leq 5$ range are 1.5 (model C) and 4
(model D).  If we now account for the mean spectral curvature of the
sources, as described in Sec.~\ref{sec:imp_curv}, and re-run the
simulations we predict $0.1$ quasars with $5 \leq z \leq 7$ (and $0.5$
with $4 \leq z \leq 5$) for model C, and now only $1$ with $5 \leq z
\leq 7$ (and $1.6$ with $4 \leq z \leq 5$) for model D.\footnote{It is worth
noting that, although the effects of spectral curvature on the derived
evolution in co-moving space densities were shown to be systematic but small in
Sec.~\ref{sec:VVmax}, they become increasingly important as the properties
of the population are extrapolated to larger redshifts.}
Such numbers represent a tiny fraction of the $\sim 1000$
flat-spectrum sources in the survey, but are consistent with a
previous estimate (Dunlop et al. 1986) that $\sim 0.5$ per cent of the
Parkes flat-spectrum sources lie at $z > 4$ (this estimate was based
on a gradual redshift cut-off very similar to the one preferred by the
analysis of Sec.~\ref{sec:cons_uncert}, but did not account for spectral
curvature).  For a no cut-off model our predicted numbers of
high-redshift quasars are an order-of-magnitude lower than the SH98
estimate of 15 quasars at $z > 5$; we believe this to be a good
illustration of the significance of the biases discussed at length in
Secs.~\ref{sec:imp_spix} and~\ref{sec:imp_curv}.

These simulations show that even restricting our
attention to two specific models, small number statistics preclude any
definitive statement on the cut-off since the probability of detecting
zero $z > 5$ objects in the no cut-off case is given by the Poisson probability
$\sim e^{-1} \approx 0.35$.  Therefore, even if the SH96
survey was extended to the whole sky, and it was certain that it
included no $z >5$ quasars, then one could only rule out model D at
roughly the 95 per cent level by using the most luminous sources.  

We conclude that the observable volume
in the high-redshift Universe is simply too small to delineate the
redshift cut-off using objects as rare as the flat-spectrum quasars
{\sl in the top decade of the RLF.} 

\subsection{Less luminous flat-spectrum sources}\label{sec:less_lum}

We next assess the chances of delineating the cut-off by analysing
redshift surveys at fainter flux-density limits.  We consider first
the GPS sources, which we argued in Sec.~\ref{sec:linking_steep} are an
important population within the top decade of the flat-spectrum RLF.
GPS sources in this luminosity regime should be the progenitors of
FRII radio sources with 151 MHz luminosities $\log_{10}(L_{151}) \sim
26.5$, so to estimate the number of GPS sources in fainter samples we
need to consider how the steep-spectrum RLF behaves at lower values of
$L_{151}$. At $z \sim 2$ the steep-spectrum RLF is constrained by the
radio source counts to flatten considerably below $\log_{10} (L_{151}) \sim
27$ (Willott et al. 1998; Willott et al. 2000b), so that lowering
the flux-density limit by say a factor of ten, would increase the
number of high-redshift GPS quasars in a given sky area by a factor of
a few rather than by orders-of-magnitude. Direct evidence in favour of
this argument comes from the GPS survey at faint ($\sim 0.05$ Jy) flux-density by Snellen et al. (1998,1999) which has found 13 GPS quasars
with a similar redshift distribution to those found in Table~\ref{tab:samples},
representing a surface density $\sim 20$ sr$^{-1}$, i.e. a factor
$\sim 4$ higher than the surface density of the most luminous sources
in the PHJFS sample.  Therefore, although the huge effort of surveying
the whole sky for GPS sources could increase the number of objects in
the high-redshift Universe considerably, and thus improve constraints
on the cut-off, the gains would be small enough that small number
statistics would remain a crucial limitation.

There is also an argument that the importance of 
GPS sources within the $\log_{10} (L_{2.7}) > 27$ population might
swiftly reverse to complete dominance by the DB population over 
the next lowest decade. It would be pure fluke if the 
FRII counterparts of the DB and GPS populations had precisely the
same characteristic value of
$\log_{10} (L_{151})$ for a given value of $\log_{10} (L_{2.7})$. If, as
seems likely (Sec.~\ref{sec:linking_steep}) the DB objects are
on average drawn from higher up the FRII RLF, then the number of these
will increase more rapidly with decreasing $L_{2.7}$
than would be the case for GPS sources.
We have presented tentative evidence that
this is actually the case in Sec.~\ref{sec:linking_steep}.
This means that studies of the
cut-off with samples dominated by DB objects, and with much less severe
problems due to small number statistics, may
be plausible at lower flux-density levels. 

Of course DP90 have made the first stab at just such an experiment,
but we have argued in Sec.~\ref{sec:top-decade} that among other problems
the sky area of their faintest sample was insufficient to add much to
constraints available from the bright sample data (P85; this paper).
The DP90 analysis also does not account for the change over from a
mixed population of GPS and DB sources in the top decade, to mainly DB
sources in the next lower decade. A new analysis that separates these
distinct populations and that takes careful account of $K-$corrections
is probably warranted, and in the longer term, a large increase in
survey area is highly desirable. The higher values of $\rho$ for the
$\log_{10} (L_{2.7}) \sim 26-27$ DB population means that the volume on our
light cone in the high-redshift Universe is sufficient to eliminate the
problem of small number statistics given a sufficiently wide-area
redshift survey. Selection of samples of flat-spectrum sources should
probably incorporate low-frequency survey data to avoid missing
objects with spectra which are intrinsically flat at $\sim$GHz
frequencies, but which appear steep because their spectra are highly
redshifted.

\section{Acknowledgements}\label{sec:ack}
This paper has benefited enormously from written comments made by
Jasper Wall and an anonymous referee, and we are very grateful to them
both. We would also like to thank Isobel Hook, Peter Shaver, Devinder Sivia,
Steve Warren and Chris Willott for helpful comments and, in the case
of Devinder, for donating some useful software. 
This research has made use of the NASA/IPAC
Extra-galactic Database, which is operated by the Jet Propulsion
Laboratory, Caltech, under contract with the National Aeronautics and
Space Administration.  MJJ thanks PPARC for the receipt of a
studentship.

\end{document}

%% file: macro.tex
\newcommand{\gtsim}{\mbox{{\raisebox{-0.4ex}{$\stackrel{>}{{\scriptstyle\sim}}
$}}}}
\newcommand{\ltsim}{\mbox{{\raisebox{-0.4ex}{$\stackrel{<}{{\scriptstyle\sim}}
$}}}}

\newcommand{\figstart}[1]
    { \begin{figure}[htb]
      \begin{picture}(0,#1) }
\newcommand{\figend}[4]
    { \end{picture}
      \special{#1}
      \caption[#2]{#3}
      \label{#4}
      \end{figure} }